\documentclass[preprint]{aastex}
\shorttitle{A spectroscopic study of I Zw 18C}
\shortauthors{Izotov et al.}
\begin{document}

\title{A spectroscopic study of component C and the extended
    emission around I Zw 18}
\author{Yuri I. Izotov}
\affil{Main Astronomical Observatory, Ukrainian National Academy of Sciences,
Golosiiv, Kyiv 03680, Ukraine}
\email{izotov@mao.kiev.ua}

\author{Frederic H. Chaffee}
\affil{W. M. Keck Observatory, 65-1120 Mamalahoa Hwy., Kamuela, HI 96743, USA}
\email{fchaffee@keck.hawaii.edu}

\author{Craig B. Foltz}
\affil{MMT Observatory, University of Arizona, 
Tucson, AZ 85721, USA}
\email{cfoltz@as.arizona.edu}

\author{Trinh X. Thuan}
\affil{Astronomy Department, University of Virginia, Charlottesville, VA 22903, 
USA}
\email{txt@virginia.edu}

\author{Richard F. Green}
\affil{National Optical Astronomy Observatories, Tucson, AZ 85726, USA}
\email{rgreen@noao.edu}

\author{Polychronis Papaderos and Klaus J. Fricke}
\affil{Universit\"ats--Sternwarte, Geismarlandstrasse 11, D--37083 G\"ottingen, 
Germany}
\email{papade@uni-sw.gwdg.de, kfricke@uni-sw.gwdg.de}

\and

\author{Natalia G. Guseva}
\affil{Main Astronomical Observatory, Ukrainian National Academy of Sciences,
Golosiiv, Kyiv 03680, Ukraine}
\email{guseva@mao.kiev.ua}

\date{Received ; accepted }



\begin{abstract}
Long-slit Keck II\footnote{W.M. Keck Observatory 
is operated as a scientific partnership among the California Institute of 
Technology, the University of California and the National Aeronautics and 
Space Administration. The Observatory was made possible by the generous 
financial support of the W.M. Keck Foundation.}, 4m Kitt Peak\footnote{Kitt 
Peak National Observatory (KPNO) 
is operated by the Association of Universities for Research in Astronomy, 
Inc., under cooperative agreement with the National Science Foundation.}, and
4.5m MMT\footnote{The MMT is a joint 
facility of the Smithsonian Institution and the University of 
Arizona.} spectrophotometric data are used to investigate the  
stellar population and the evolutionary status of I Zw 18C, the 
faint C component of the nearby blue compact dwarf galaxy I\ Zw\ 18. 
Hydrogen H$\alpha$ and H$\beta$ emission lines are detected in the spectra
of I Zw 18C, implying that ionizing massive stars are present.
High signal-to-noise Keck II spectra of different
regions in I Zw 18C reveal H$\gamma$, H$\delta$ 
and higher order hydrogen lines in absorption.
Several techniques are used to constrain the age of the stellar
population in I Zw 18C. Ages derived from two different 
methods, one based on the equivalent widths of the 
H$\alpha$, H$\beta$ emission lines and the other on H$\gamma$, H$\delta$ 
absorption lines are consistent with a 15 Myr instantaneous burst
model. We find that a small 
extinction in the range $A_V$ = 0.20 -- 0.65 mag is needed to
fit the observed spectral energy distribution of I Zw 18C with that
model. In the case of constant star formation, all
observed properties are consistent with stars forming continuously between
$\sim$ 10 Myr and $\la$ 100 Myr ago. We use all available observational 
constraints for I Zw 18C, including those obtained from 
{\sl Hubble Space Telescope} color-magnitude diagrams, to argue that
the distance to I Zw 18 should be as high as $\sim$ 15 Mpc.
The deep spectra also reveal extended ionized gas emission around I Zw 18.
H$\alpha$ emission is detected as far as 30\arcsec\ from it.
To a $B$ surface brightness limit of $\sim$ 27 mag arcsec$^{-2}$
we find no observational evidence for extended
stellar emission in the outermost regions, at distances $\ga$ 15\arcsec\ 
from I Zw 18. 
\end{abstract}


\section{Introduction}
%
I Zw 18 remains the most metal-poor blue compact dwarf (BCD) galaxy known
since its discovery by Sargent \& Searle (1970).
Later spectroscopic observations by Searle \& Sargent (1972), Lequeux et al. 
(1979), French (1980), Kinman \& Davidson
(1981), Pagel et al. (1992), Skillman \& Kennicutt (1993), Martin (1996),
Izotov \& Thuan (1998),
V\'ilchez \& Iglesias-P\'aramo (1998), Izotov \& Thuan (1999) and Izotov et al.
(1999) have confirmed its low metallicity with an oxygen abundance of only 
$\sim$ 1/50 the solar value.

Zwicky (1966) described I Zw 18 as a double system of compact galaxies,
which are in fact two bright knots of star formation with an angular 
separation of 5\farcs8. 
These two star-forming regions (Fig. \ref{Fig1}) are referred to as the 
brighter northwest (NW) and fainter southeast (SE) components.
Later studies by Davidson, Kinman \& Friedman (1989) and Dufour \&
Hester (1990) have revealed a more complex optical morphology 
with additional diffuse features.
The most prominent diffuse feature, hereafter I Zw 18C (Fig. \ref{Fig1}), 
is a blue irregular star-forming region $\sim$ 22\arcsec\ northwest of the NW 
component. Dufour, Esteban \& Casta\~neda (1996a), Izotov \& Thuan (1998)
and van Zee et al. (1998) have shown I Zw 18C
to have a systemic radial velocity equal to that of the ionized gas in the 
NW and SE components, thus establishing its physical association to I\ Zw\ 18. 
Furthermore, van Zee et al. (1998) have shown that I Zw 18C is
embedded in a common H {\sc i} envelope with the NW and SE components.

Many studies have been focused on the evolutionary state of
I Zw 18. Searle \& Sargent (1972) and Hunter \& Thronson (1995) 
have suggested that it may be a young galaxy,
recently undergoing its first burst of star formation. The latter authors 
concluded from {\sl Hubble Space Telescope} ({\sl HST}) 
images that the colors of 
the diffuse unresolved component surrounding the SE and NW  regions are 
consistent with a population of B and early A stars, i.e. with no evidence for 
older stars. Ongoing massive star formation in I\ Zw\ 18 is 
implied by the discovery of a population of Wolf-Rayet stars in the NW 
component (Izotov et al. 1997a; Legrand et al. 1997). 

From the analysis of color-magnitude diagram (CMD) based on {\sl HST}
images, Dufour et al. (1996b) concluded that star formation 
in I Zw 18 began at least 30 -- 50 Myr ago and continuing to the present. 
Martin (1996) and Dufour et al. (1996b) have discussed the
properties of expanding superbubbles of ionized gas driven by supernova
explosions and have inferred dynamical ages of 15 -- 27 Myr and 
13 -- 15 Myr respectively. 

    Recently, Aloisi, Tosi \& Greggio (1999) have discussed the star formation
history in I Zw 18 using the same {\sl HST} WFPC2 archival data
(i.e. those by Hunter \& Thronson (1995) and Dufour et al. (1996b)). 
They compared observed CMDs and luminosity functions with synthetic ones and
concluded that there were two episodes of star formation in I Zw 18,
the first one occuring over the last 0.5 -- 1 Gyr, an age more than 10 
times larger than that derived by Dufour et al. (1996b), and the second one 
with more intense activity taking place between 15 and 20 Myr ago. 
No star formation has occurred within the last 15 Myr. 
\"Ostlin (2000) from {\sl HST} NICMOS $J$ and $H$ observations concluded that
a 1 -- 5 Gyr old stellar population is present.

The component I Zw 18C has not been studied in such detail
mainly because of its faintness. Its flux-calibrated optical 
spectra (Izotov \& 
Thuan 1998; van Zee et al. 1998) reveal a blue continuum with weak Balmer 
absorption features and faint H$\alpha$ and H$\beta$ in emission. Such spectral 
features suggest that the H {\sc ii} region is probably ionized by 
a population of early B stars.

Dufour et al. (1996b) found in a $(B-V)$ vs. $V$ CMD analysis 
of I Zw 18C a well-defined upper 
stellar main sequence indicating an age of the blue stars of $\sim$ 40 Myr. 
However, numerous faint red stars were also present in the CMD, 
implying an age of 100 -- 300 Myr. Those authors concluded that I Zw 18C
consists of an older stellar population with an age of several hundred Myr, 
but which has experienced recently a modest starburst in its southeastern half
as evidenced by the presence of blue stars and H$\alpha$ emission. 
Aloisi et al. (1999) estimated an age for I Zw 18C not exceeding 0.2 Gyr. 

We use here high signal-to-noise 4.5m MMT, 4m KPNO and Keck II spectroscopy
to study the evolutionary status of I Zw 18C. We also discuss the nature
of the extended emission in the outermost regions of I Zw 18. Our motivation for
this study is the following. Until now, age estimates for the stellar 
populations in I Zw 18C are based solely on {\sl HST}
CMDs. While in principle CMD studies are powerful tools for studying
stellar populations, they critically depend on the adopted distance to
the galaxy and the interstellar extinction, which are a priori unknown. 
We use here distance-independent techniques
based on spectroscopic observations to derive the age of the stellar 
populations in I Zw 18C.

Concerning the outermost regions of I Zw 18, while
gaseous emission is an important contributor to the total light
in the vicinity of the star-forming regions, there was no evidence for
extended stellar emission at distances as far as 20\arcsec\ from 
the H {\sc ii} regions (e.g., Dufour et al. 1996b; Izotov et
al. 1999). However, in some recent papers (e.g., Legrand 2000; Legrand et al.
2000; Kunth \& \"Ostlin 2000) such an old extended stellar population 
has been postulated. We use deep MMT and Keck II spectroscopic observations
to clarify the origin of the extended emission and estimate the
contribution of the ionized gas to it.

The observations and data reduction are described in Sect. 2.
The properties of the stellar population in I Zw 18C are discussed in Sect. 3.
In Sect. 4 we discuss the properties of the outlying regions of I Zw 18.
Our results are summarized in Sect. 5.

\section{Observations and data reduction}

 The Keck II spectroscopic observations of I Zw 18C were carried out
on January 9, 2000 with  the low-resolution imaging spectrograph (LRIS)
(Oke et al. 1995), using the
300 groove mm$^{-1}$ grating which provides a dispersion 
2.52 \AA\ pixel$^{-1}$ and a
spectral resolution of about 8 \AA\ in first order. 
The slit was 
1\arcsec$\times$180\arcsec, centered on the brightest central region 
(region C) of I Zw 18C and oriented with 
a position angle P.A. = --80$^{\circ}$ (slit orientation ``1'' in
Fig. \ref{Fig1}). No binning along the spatial axis has been done, yielding a 
spatial sampling of 0\farcs2 pixel$^{-1}$. The total exposure time was 60 min,
broken into three 20 min exposures. All exposures were taken at airmass of
1.42. The seeing was 0\farcs9.

MMT spectroscopic observations of I\ Zw\ 18 and I Zw 18C were carried out in 
the nights of 1997 April 29 and 30. 
A signal-to-noise ratio S/N $\ga$ 50 was reached in the continuum
of the bright NW region of I Zw 18. 
Observations were made in the blue channel of the MMT spectrograph 
using a highly optimized Loral 3072 $\times$ 1024 CCD detector. 
A 1\farcs5 $\times$ 180\arcsec\ slit was used along with a 300 groove mm$^{-1}$
grating in first order and an L-38 second-order blocking filter.
This yields a spatial resolution along the slit of 0\farcs3 pixel$^{-1}$,
a scale perpendicular to the slit of 1.9 \AA\ pixel$^{-1}$, a spectral
range 3600 -- 7500 \AA, and a spectral resolution of $\sim$ 7 \AA\ (FWHM). To
improve the signal-to-noise ratio, CCD rows were binned by a factor of 2, 
yielding a final spatial sampling of 0\farcs6 pixel$^{-1}$. 
The total exposure time was 180 minutes broken up in six 
subexposures, 30 minutes each. 
All exposures were taken at airmasses $\la$ 1.1 -- 1.2.
The seeing during the observations was 0\farcs7 FWHM. 
The slit was oriented in a position angle P.A. = --41$^{\circ}$ to permit 
observations of the NW and SE regions in I Zw 18 
and the eastern region (region E) in I Zw 18C simultaneously 
(slit orientation ``2'' in Fig. \ref{Fig1}). 

   The Kitt Peak 4m observations have been obtained 
on 18 March 1994 with the Ritchey-Chr\'{e}tien RC2 spectrograph used
in conjunction with the T2KB 2048$\times$2048 CCD detector. We use a 
2\arcsec$\times$300\arcsec\ slit with the 
KPC--10A grating (316 lines mm$^{-1}$) in first order, with a GG 385 order 
separation filter. This filter cuts off all second-order contamination for 
wavelengths blueward of 7400\AA\, which is the wavelength region of interest 
here. The above instrumental set-up gave a spatial scale
along the slit of 0.69 arcsec pixel$^{-1}$, a scale perpendicular to the slit
of 2.7\AA\ pixel$^{-1}$, a spectral range of 3500--7500\AA\ and a spectral
resolution of $\sim$ 5\AA. All exposures were taken at airmass of 1.1.
The seeing was 1\farcs5. 
The slit was oriented along the SE -- NW direction at a position angle 
of --41$^\circ$, the same as that during the MMT observations 
(slit orientation ``2'' in Fig. \ref{Fig1}). 
The total exposure time was 60 minutes and was broken up into 3 subexposures.

The spectrophotometric standard stars Feige 34 and HZ 44 were observed for 
flux calibration during each of three sets of the observations. 
Spectra of Hg-Ne-Ar (Keck II) and He-Ne-Ar (MMT and 4m 
KPNO) comparison lamps were obtained before and after each
observation to provide the wavelength calibration.

Data reduction of spectral observations was carried out using the IRAF 
software package.\footnote{IRAF: the Image Reduction
and Analysis Facility is distributed by the National Optical Astronomy 
Observatory, which is operated by the Association of Universities for
Research in Astronomy, In. (AURA) under cooperative agreement with the
National Science Foundation (NSF).} 
This included bias subtraction, cosmic-ray removal
and flat-field correction using exposures of a quartz incandescent
lamp. After wavelength calibration, night-sky background subtraction, 
and correcting for atmospheric extinction, each frame was
calibrated to absolute fluxes. One-dimensional spectra were
extracted by summing, without weighting, different numbers
of rows along the slit depending on the exact region of interest.
We have extracted spectra of two regions in I Zw 18C (Fig. \ref{Fig1}): 
(1) the brightest region C (Keck II observations)
and (2) the eastern region E (all observations). 
The extracted spectra are shown in Fig. \ref{Fig2}. 
Additionally, spectra of outlying regions of I Zw 18 
at different distances from it have been extracted.

\section{The stellar population in I Zw 18C}

One of the key problems discussed over the last three decades is the 
evolutionary status of I Zw 18: is this galaxy young or old? 
The evolutionary status of I Zw 18C has not been discussed in comparable detail.
High signal-to-noise spectra of I Zw 18C reveal blue continua and show
only emission and absorption hydrogen Balmer 
lines. 
Heavy element emission lines are not detected in the spectra, which precludes 
a metallicity determination of I Zw 18C.
For the sake of simplicity we assume the heavy element mass fraction in 
I Zw 18C to be $Z_\odot$/50, the same value as in I Zw 18. 
However, the spectra obtained for I Zw 18C allow to study stellar 
populations and constrain their age with various techniques.

\subsection{Age determination\label{age}}
%

A useful technique for determining the age of a galaxy is to fit its observed 
spectral energy  distribution (SED) by theoretical SEDs calculated 
for various stellar population ages and star formation histories. This 
method (alone or in combination with photometric data) has been applied to 
some extremely metal-deficient BCDs with $Z$ = (1/20 -- 1/40)$Z_\odot$ 
(e.g., SBS 0335--052 (Izotov et al. 1997b; Papaderos et al. 1998), 
SBS 1415+437 (Thuan et al. 1999a), Tol 1214--277 (Fricke et al. 2001)).
It was shown that, after subtraction of ionized gas emission, the underlying 
stellar components of these galaxies are consistent with populations not 
older than a few hundred Myr.

However, the spectral energy distribution fitting method is
subject to uncertainties in the extinction, resulting in an age overestimate,
if the adopted extinction is too low. Therefore, other methods are desirable
to constrain the stellar population ages. We discuss in this Section two 
such methods, one relying on the Balmer nebular emission line equivalent 
widths and the other on the Balmer stellar absorption line equivalent widths.  

\subsubsection{Age from the nebular emission lines \label{ageem}} 

 On the assumption of a dust-free ionization-bounded H {\sc ii} region, the 
strongest hydrogen recombination emission lines H$\alpha$ and H$\beta$ provide 
an estimate of the age of the young stellar population when O and early B stars 
are still present. However, even if dust is present in H {\sc ii} regions,
the age estimate is quite robust. This is because the ionizing flux from such 
a young stellar population and hence the equivalent widths of the Balmer 
emission lines have a very strong temporal evolution. 
Therefore, the dating method
based on the H$\alpha$ and H$\beta$ emission lines is relatively insensitive
to dust extinction.

The H$\alpha$ and H$\beta$ emission lines are 
detected in I Zw 18C in both regions C and E (Fig. \ref{Fig2}).
Their fluxes and equivalent widths are listed in Table \ref{Tab1}.
The exception is the Keck II spectrum of region E, where H$\beta$ emission
was not detected. This non-detection is probably due to the patchy 
distribution of the ionized gas. In Fig.~\ref{Fig3}a we compare the 
measured H$\alpha$ and H$\beta$ emission line equivalent widths with those 
predicted for an instantaneous burst as a function of age. The theoretical 
dependences have been kindly calculated for us by D. Schaerer using 
the Schaerer \& Vacca (1998) code with the $Z$ = 0.0004
Geneva evolutionary tracks from Lejeune \& Schaerer (2001). They are shown
by solid lines. The age derived from 
different hydrogen nebular emission lines in the various spectra
is in a narrow range around $\sim$ 15 Myr. Hence, the gas in I Zw 18C is 
likely to be ionized by early B stars.

\subsubsection{Age from the hydrogen stellar absorption lines\label{ageab}}

   Another method of stellar population age determination is based on the 
equivalent widths of absorption features. This method probes larger ages 
as compared to the previous method because the most prominent 
hydrogen absorption lines are formed in the longer-lived A stars.

 Gonzalez Delgado \& Leitherer (1999) and Gonzalez Delgado,
Leitherer \& Heckman (1999) have calculated a grid
of synthetic profiles of stellar hydrogen Balmer absorption lines for 
effective temperatures and gravities, characteristics of galaxies with 
active star formation.
They developed evolutionary stellar population synthesis models, synthesizing 
the profiles of the hydrogen Balmer absorption lines from H$\beta$ to H13 
for an instantaneous burst with an age ranging from $10^6$ to $10^9$ yr.  
The calculations were made for a stellar initial mass 
function with Salpeter slope and with mass cutoffs 
$M_{\rm low}$ = 1 $M_{\odot}$ and $M_{\rm up}$ = 80 $M_{\odot}$.

 The H$\gamma$, H$\delta$ and higher order hydrogen absorption lines 
due the underlying stellar populations 
are clearly seen in the Keck II spectra of I Zw 18C (Fig.
\ref{Fig2}a - \ref{Fig2}b). Some hydrogen absorption lines are also seen
in the 4m KPNO and MMT spectra (Fig. \ref{Fig2}c - \ref{Fig2}d).
However, the signal-to-noise ratio is not high enough in the latter two
spectra to measure equivalent widths.
Although higher-order hydrogen Balmer absorption lines are 
seen in the spectrum of I Zw 18C, they are not suitable for 
age determination because of (a) the relatively low signal-to-noise ratio 
at short wavelengths and uncertainties in the placement of the continuum in the 
blue region caused by many blended absorption features, and (b) the weak 
dependence of their equivalent widths on age (Gonzalez Delgado et al. 1999).

  In Table \ref{Tab2} we show the equivalent widths of the H$\delta$
and H$\gamma$ absorption lines measured in the spectra of regions C and E
in I Zw 18C. We need to correct the equivalent widths
of the absorption lines for the contamination by nebular emission. 
For region C we use the intensity of the H$\beta$ emission line to 
calculate the intensity of the H$\delta$ emission line
adopting case B at the electron temperature of $\sim$
20000K (e.g., Aller 1984). We do not use the H$\gamma$ absorption line in the
spectrum of this region because of the strong contamination by
nebular emission. The H$\beta$ emission line is not definitely
detected in the Keck II spectrum of region E. For this region, the intensity of 
H$\alpha$ emission line is used to correct equivalent widths of the absorption
lines for the same effect. The corrected equivalent widths of H$\delta$ and
H$\gamma$ absorption lines are shown in Table \ref{Tab2}.

 In Fig.~\ref{Fig3}b we show by solid lines the predicted behaviour of the 
equivalent widths of the H$\gamma$ and H$\delta$ absorption lines with age for 
an instantaneous burst at a metallicity $Z$ = 1/20 $Z_\odot$ 
(Gonzalez Delgado et al. 1999). The measured equivalent widths are shown
for region C by filled circles and for region E by stars. Their
values are consistent with an age of $\sim$ 15 Myr. This 
age estimation is in excellent agreement  with that obtained from the nebular 
emission line analysis implying that the light of I Zw 18C is dominated by a
young stellar population. However, the age we derive here is significantly 
lower than the value of 40 Myr derived by Dufour et al. (1996b) for the 
brightest stars from {\sl HST} CMDs of I Zw 18C.

\subsubsection{Uncertainties}

Although our two age estimates are consistent with each other, there are
a number of uncertainties which may affect the result.

A major uncertainty is the unknown metallicity of the stars in
I Zw 18C. We have assumed for simplicity the metallicity to be equal to that
of the ionized gas in I Zw 18. Note, however, that a lower stellar 
metallicity would increase the age and vice versa.

To estimate an age from the emission lines we have assumed a dust-free
ionization-bounded H {\sc ii} region. If the H {\sc ii} region is 
density-bounded or dust is present, then some of the ionizing photons escape 
the H {\sc ii} region or are absorbed. Equivalent widths of the Balmer emission 
lines give in this case an upper limit to the age.

Another source of uncertainty comes from the small number of 
massive stars in I Zw 18C. Our age estimates are based
on models where the stellar initial mass function is well-behaved and
can be approximated by an analytical function. However, small number statistics 
can introduce stochastic fluctuations at the high star mass end of the IMF.
Recently Cervi\~no, Luridiana \& Castander (2000) have analyzed how such
stochastic effects influence the observed parameters of young stellar
clusters with solar metallicity such as the H$\beta$ equivalent width and the 
number of the ionizing photons. The number
of ionizing photons in I Zw 18C derived from the H$\beta$ emission line 
(Table 1) is $\sim$ 2 $\times$ 10$^{48}$ s$^{-1}$ and 4 $\times$ 
10$^{48}$ s$^{-1}$ for the E and C regions respectively.
With an equivalent width $\sim$ 6 -- 7 \AA\ of the H$\beta$ emission line, this
corresponds to the case of a 10$^3$ $M_\odot$ stellar cluster 
(Cervi\~no et al. 2000). For such a cluster the age variations at a fixed 
H$\beta$ equivalent width can be as high as 15 percent at the 90\% 
confidence level. Hence, the age of I Zw 18C derived from the emission
lines can lie in the range $\sim$ 10 -- 25 Myr, with a central value of
15 Myr. Similarly, age estimates based on absorption lines can also be slightly
modified by stochastic effects. However, calculations are not yet available in
the literature.

Gonzalez Delgado et al. (1999) do not calculate the temporal evolution of
the equivalent widths of the Balmer absorption lines for the heavy element
mass fraction $Z$ = $Z_\odot$/50. Therefore, we use models with
$Z$ = $Z_\odot$/20. Extrapolation to the metallicity of I Zw 18 would result
in a $\la$ 1\AA\ decrease of the equivalent widths at a fixed age, 
or an age increase of up to 25 Myr.

Finally, the age determination depends on the star formation history in 
the galaxy which we consider next. 

\subsubsection{Continuous star formation\label{extend}}

  Our estimates for the stellar population age in I Zw 18C in Sect. 
\ref{ageem} and \ref{ageab} are based on the assumption of an instantaneous 
burst of star formation. Now we discuss how that age changes if continuous star 
formation is considered. We adopt a constant star formation rate in the interval 
between the initial time $t_{\rm i}$ when star formation starts and 
the final time $t_{\rm f}$ when it stops. Time is zero at the present epoch
and increases into the past. 

Using model equivalent width of the emission and absorption lines and spectral 
energy distributions for instantaneous bursts (Schaerer, private communication;
Lejeune \& Schaerer 2001; Gonzalez Delgado et al. 1999), we calculate the 
temporal evolution of the hydrogen emission and absorption line equivalent
widths for continuous star formation. The results
of calculations are presented in Figure \ref{Fig3}. By dashed,
dot-dashed and dotted lines are shown the temporal dependences of the equivalent
widths of the H$\beta$ and H$\alpha$ emission lines (Fig. \ref{Fig3}a), and of 
the H$\delta$ and H$\gamma$ absorption lines (Fig. \ref{Fig3}b) for continuous 
star formation starting at time $t_{\rm i}$, as defined by the abscissa value, 
and stopping at $t_{\rm f}$ = 5, 8 and 12.5 Myr, respectively. 
In other words, the equivalent widths of the above four lines
in the spectrum of the stellar population formed between $t_{\rm i}$ and 
$t_{\rm f}$ have a value $EW$ at time $t_{\rm i}$ in Fig. \ref{Fig3}a and 
\ref{Fig3}b. At a fixed $EW$, the general trend seen from Fig. \ref{Fig3} for
continuous star formation is that the younger the youngest stars,
the larger the time interval $t_{\rm i}-t_{\rm f}$, and the older the oldest 
stars. Another feature is that, at a fixed age $t_{\rm f}$ of the youngest 
stars, the age $t_{\rm i}$ of the oldest stars derived from the observed 
emission line equivalent widths, differs from that derived from the observed 
absorption line equivalent widths. 
In particular, in the model where star formation stopped 5 Myr
ago (dashed lines), the age of the oldest stars derived from hydrogen emission 
lines exceeds 100 Myr, while the age of the oldest stars  derived from 
hydrogen absorption lines is only $\sim$ 50 Myr. This model seems
to be excluded by consideration of the luminosity of the ionizing
radiation. The most massive stars in the stellar population with age
5 Myr would have masses as high as 40 $M_\odot$ (Meynet et al. 1994). 
The number of the ionizing
photons produced by a single 40 $M_\odot$ star is equal to $N$(Lyc) 
$\approx$ 1.5 $\times$ 10$^{49}$ s$^{-1}$ (Vacca, Garmany \& Shull 1996),
larger than that derived from the observed flux of the H$\beta$ emission line 
in I Zw 18C (Table \ref{Tab1}), assuming an ionization-bounded H {\sc ii}
region. There can be an upward correction factor of $\la$ 2 
due to the extinction, but the corrected $N$(Lyc) would still be below 
the value for a single 40 $M_\odot$ star. These estimates can however
be modified by massive star small number statistics 
caused by the stochastic nature of star formation.

Though smaller, the difference between the age of the oldest stars derived
from the equivalent widths of emission lines (50 Myr) and that derived
from the equivalent widths of absorption lines (40 Myr), is present in the
continuous star formation model with age of the youngest stars equal to 8 Myr 
(dot-dashed
lines in Fig. \ref{Fig3}). However, this difference is small in the continuous
star formation model which stopped 12.5 Myr ago (dotted lines). In this
model, the age of the oldest stars should be $\sim$ 25 Myr to consistently
explain the observed hydrogen line equivalent widths.

Hence, similarly to the case of an instantaneous burst, we conclude that the 
observations of I Zw 18C are best reproduced by a short star formation episode 
which occurred continuously between $\sim$ 10 Myr and $\sim$ 25 Myr ago.
Uncertainties in the observations and models may extend this range to between
$\sim$ 10 Myr and $\la$ 100 Myr ago.

%
\subsection{Synthetic spectral energy distribution}
%

 A useful constraint on the stellar population age can be obtained from
the spectral energy distribution. This method, as already noted,
is subject to interstellar extinction. However, when used in conjunction with 
the methods discussed in Sect. \ref{age} it provides a powerful tool for 
studying stellar populations by allowing to derive simultaneously the age and 
the extinction of the same region.

 To fit the observed spectral energy distributions we use model SEDs 
calculated by D. Schaerer using the Schaerer \& Vacca
(1998) code and the $Z$ = 0.0004 Geneva evolutionary tracks of 
Lejeune \& Schaerer (2001). The contribution of the ionized gas was also
included. This contribution is small
because the equivalent widths of hydrogen emission lines in I Zw 18C are low.

Because the observed spectral energy distribution is extinction-dependent, 
the extinction can be obtained for regions with known ages as derived from the 
equivalent widths of the hydrogen emission and absorption lines. We consider 
the case of the 15 Myr instantaneous burst stellar population 
discussed in Sect. \ref{age}. First assume $C$(H$\beta$) = 0,  where
$C$(H$\beta$) = $E$($B-V$)/1.47 (Aller 1984). 
Comparison of the observed Keck II spectra of regions C and E
in I Zw 18C with the theoretical SEDs (bottom spectra in Fig. \ref{Fig4})
shows that theoretical SEDs are bluer than the observed 
extinction-uncorrected spectra. Evidently, interstellar extinction is present
in I Zw 18C and it modifies the observed SED. We derive $C$(H$\beta$) = 0.3
for region C and $C$(H$\beta$) = 0.1 for region E to achieve the best agreement 
between the extinction-corrected observed SEDs and the theoretical SEDs
(upper spectra in Fig. \ref{Fig4}). For comparison, we show 
in Fig. \ref{Fig4}a by a dotted line the theoretical SED for a 40 Myr stellar 
population which does not provide as good a fit.

A theoretical 15 Myr stellar population SED also fits well the 4m KPNO and MMT
spectra of region E extinction-corrected for $C$(H$\beta$) = 0.1
(Fig. \ref{Fig5}). The theoretical 40 Myr stellar population SED with 
$C$(H$\beta$) = 0.1 fits less well (bottom solid line).

Some support for a larger extinction in region C than in region E comes from the
observed H$\alpha$-to-H$\beta$ flux ratios (Table \ref{Tab1}). 
They are respectively equal to 4.5 and 3.8, corresponding to $C$(H$\beta$) 
$\sim$ 0.7 and 0.4. However, correction for underlying stellar absorption
results in lower extinction coefficients.

We note that we have not corrected the observed SEDs for the effect of 
atmospheric refraction. If such an effect were to be 
important, it can selectively 
decrease the blue light relatively to the red light, leading us 
to derive erroneously high extinction for regions C and E. 
Indeed one may suspect that such an effect would be important for 
the Keck II spectrum which was 
obtained with a narrow slit of 1\arcsec\ at an airmass of 1.4. 
Filippenko (1982) has shown that atmospheric 
dispersion can produce an offset as high as $\sim$ 1\farcs2 of the blue region
near [O {\sc ii}] $\lambda$3727 relative to the red region near H$\alpha$
$\lambda$6563 at this airmass. However, his
calculations have been done for an altitude of 2 km, while the Keck II spectrum
was obtained at an altitude about twice that. Furthermore, emission from the C
component is extended and originates in a region significantly larger
than the width of the slit, reducing the effect of the atmospheric dispersion.
Perhaps the best argument for such an effect not to be important comes  
from the comparison of our different spectra of the same region. 
Although the Keck, MMT and 4m spectra of region E were obtained with
different slit widths at different airmasses, they are all well 
fitted by the same 15 Myr single stellar population model.

We have thus reached two important conclusions for I Zw 18C: (1) the stellar 
population responsible for its observed SED is very young, with an age of 
$\sim$ 15 Myr and (2) the region is characterized by a varying interstellar 
extinction implying the presence of non-uniformly
distributed absorbing material. I Zw 18C is not the only very metal-deficient
object to have a clumpy dust distribution. Earlier similar
conclusions have been reached for the metal-deficient BCD SBS 0335--052
by Izotov et al. (1997b, 1999) and Thuan et al. 
(1997, 1999b). While the brightest and youngest star-forming region in 
SBS 0335--052 is relatively free of dust, extinction is higher at the
location of the fainter and older super star clusters. Clumpy 
regions with large extinctions are clearly seen in the
{\sl HST} $V-I$ image of SBS 0335--052 (Thuan et al. 1997).

\subsection{Photometric constraints}

  We have arrived at the conclusion that the
light from I Zw 18C is dominated by stars $\sim$ 15 Myr old. Is this conclusion 
consistent with the photometric data? In this section, we compare the predicted 
colors for the young stellar population with integrated broad-band colors of 
I Zw 18C obtained from ground-based and 
{\sl HST} photometric observations. We also discuss the consistency between
the properties of the stellar population in I Zw 18C obtained from the 
spectroscopic data with those obtained from analysis of {\sl HST} WFPC2 CMDs.

\subsubsection{Ground-based and HST broad-band photometry}

In Table \ref{Tab3} we show the observed integrated $V$ magnitude and colors 
of I Zw 18C. The second column shows these quantities without correction for 
interstellar extinction. Because the spectroscopic data imply 
the presence of extinction in I Zw 18C, we also show in the third 
column the colors corrected for interstellar extinction with 
$C$(H$\beta$) = 0.3 or $A_V$ = 0.65 mag. These values are for the brightest 
region C. The extinction is lower in the fainter region E. The faint 
northwestern region of I Zw 18C appears to be redder as compared to other 
regions (Dufour et al. 1996b), but the lack of spectroscopic data prevents
us from determining the interstellar extinction in that region. We assume 
that the extinction derived for region C to be representative for the whole 
galaxy. 

We compare the observed integrated colors of I Zw 18C with those predicted
by instantaneous burst models for different ages. The first set
of models shown in Table \ref{Tab3} is the same as the one used 
for fitting the SEDs with an heavy element mass fraction
$Z$ = $Z_\odot$/50 and based on Geneva stellar evolutionary tracks.
Another set of predicted colors based on the Padua stellar evolution
models has been calculated by Tantalo et al. (1996) for a single stellar
population and a heavy element mass fraction of $Z_\odot$/50.
Comparison of the two sets of models shows that colors based on the
Padua stellar evolution models are systematically redder at a fixed age
as compared to those based on the Geneva ones. Consequently, the use of Padua 
models results in younger ages as compared to Geneva models.

In the following we compare the observed colors to the modeled ones based on
Geneva tracks.
It is seen from Table \ref{Tab3} that the colors uncorrected for 
extinction are well reproduced by the model with a 100 Myr stellar population.
However, with this age the predicted equivalent widths of the 
hydrogen emission lines are too small ($EW$(H$\beta$) $\la$ 0.1\AA,
$EW$(H$\alpha$) $\la$ 0.3\AA) as compared to the observed ones 
(Fig. \ref{Fig3}a). On the other hand, the predicted equivalent widths of 
the hydrogen absorption lines ($EW$(H$\delta$) $\ga$ 10\AA, $EW$(H$\gamma$) 
$\ga$ 8\AA) are too high (Fig. \ref{Fig3}b). 
Again, to put observations into agreement with models, interstellar extinction 
has to be invoked. Indeed, all observed colors corrected for an extinction with
$C$(H$\beta$) = 0.3 are in fair agreement with predicted ones for a 15 -- 20
Myr single stellar population.

 Our conclusions do not change significantly in the case of continuous
star formation. In Fig. \ref{Fig6}a -- \ref{Fig6}c we show by solid lines
the theoretical dependences on age of the ($U-B$), ($B-V$) and ($V-I$)
colors in the case of constant continuous star formation, 
for different choices of $t_{\rm i}$ and $t_{\rm f}$. The observed
colors uncorrected for extinction (dashed lines) can be fitted by models with 
star formation starting at $t_{\rm i}$ = 100 -- 300 Myr. However, these
models predict too low an equivalent width for the 
 H$\alpha$ emission line and too large an equivalent width for the H$\delta$
absorption line (Fig. \ref{Fig6}d -- \ref{Fig6}e). Furthermore, models with
star formation stopping at $t_{\rm f}$ $\ga$ 40 Myr are excluded for the whole 
range of $t_{\rm i}$ (Fig. \ref{Fig6}d -- \ref{Fig6}e). To have the observed
colors come into agreement with the observed equivalent widths of 
the Balmer lines, a non-negligible extinction must be assumed. 
We show by dotted lines in Fig. \ref{Fig6}a -- \ref{Fig6}c 
the extinction-corrected colors with
two values of the reddening, $E$($B-V$) = 0.1 and 0.15. In the latter case, the
colors are explained by models with constant star formation starting
at an age $t_{\rm i}$ $\sim$ 30 -- 100 Myr (filled and open circles) and 
stopping at an age $t_{\rm f}$ = 8 -- 12 Myr. Observational uncertainties
will only slightly increase this age range.

We conclude that our broad-band photometric data are consistent with a young
stellar population and a non-negligible interstellar extinction in I Zw 18C.
We emphasize that the ages derived above hold only for the brightest regions
of the C component. We cannot exclude the possibility that
the age of the stellar population in regions of I Zw 18C,
not covered by our spectroscopic observations, may be larger.

\subsubsection{Stellar color-magnitude diagrams}

CMD analysis is a powerful tool for studying stellar populations. However,
as already pointed out,  
this method is sensitive to the adopted extinction and distance. While
the extinction can be derived from spectroscopic observations,
the determination of the distance is more uncertain.

 A distance of $\sim$ 10 Mpc to I Zw 18 has generally been adopted for 
analyzing the CMDs (Hunter \& Thronson 1995, 
Dufour et al. 1996b and Aloisi et al. 1999).
This assumes that the observed heliocentric radial velocity of the galaxy 
$\sim$ 740 km s$^{-1}$ is a pure Hubble flow velocity, and a Hubble constant
$H_0$ = 75 km s$^{-1}$ Mpc$^{-1}$. Adopting this distance would lead to a
conflict with the well-observed ionization state of I Zw 18C.
At 10 Mpc the brightest stars observed in I Zw 18C would
have absolute $V$ magnitudes fainter than --6 mag (Dufour et al. 1996b; 
Aloisi et al. 1999). In that case, comparison with evolutionary tracks 
implies that the most massive stars 
in I Zw 18C (called the C component by Dufour et al. (1996b)) would 
have masses less than 
9 $M_\odot$. The age of the stellar population with such an upper mass limit 
is at least 40 Myr (e.g., Dufour et al. 1996b), larger than the one
derived from the equivalent widths of hydrogen emission and absorption lines
($\sim$ 10 -- 25 Myr).
If the upper stellar mass limit of 9 $M_\odot$ derived by 
Dufour et al. (1996b) and Aloisi et al. (1999) for I Zw 18C is
correct, then ionized gas should not be present in it because of
the absence of early B stars. 
But H$\alpha$ and H$\beta$ are clearly observed. 
Our derived age of $\sim$ 15 Myr implies that early B 
stars with masses as high as $\sim$ 15 $M_\odot$ are present in 
I Zw 18C, if an instantaneous burst of star formation is assumed. 
In the case of continuous star formation, the age of the youngest
stars would be smaller and the upper mass 
limit larger to account for the presence of the ionized gas. 
We argue therefore that the stellar absolute magnitudes derived by 
Dufour et al. (1996b) and Aloisi et al. (1999) from their CMDs are too faint 
because they are based on too small an adopted distance. \"Ostlin (2000)
assumed a distance of 12.6 Mpc to analyze 
his {\sl HST} NICMOS CMD. However, even 
this distance is not enough to explain the ionization state of I Zw 18C.

An additional effect is due to extinction, with $A_V$ = 0.65 mag for the 
region C. Correcting for extinction and increasing the distance by a factor of 
$\sim$ 1.5 to $\sim$ 15 Mpc would make the most massive stars more luminous by 
a factor of $\sim$ 4 and push the mass upper limit to $\sim$ 15 $M_\odot$. 
A stellar population with such an upper mass limit would provide enough ionizing 
photons to account for the observed emission lines in I Zw 18C. 
Furthermore, the age of the brightest stars in the CMDs of I Zw 18C would be
$\sim$ 15 Myr, consistent with that derived from the hydrogen emission and 
absorption lines.

\section{The extended emission in I Zw 18}

\begin{figure}
\epsscale{0.7}
\figurenum{1}
\plotone{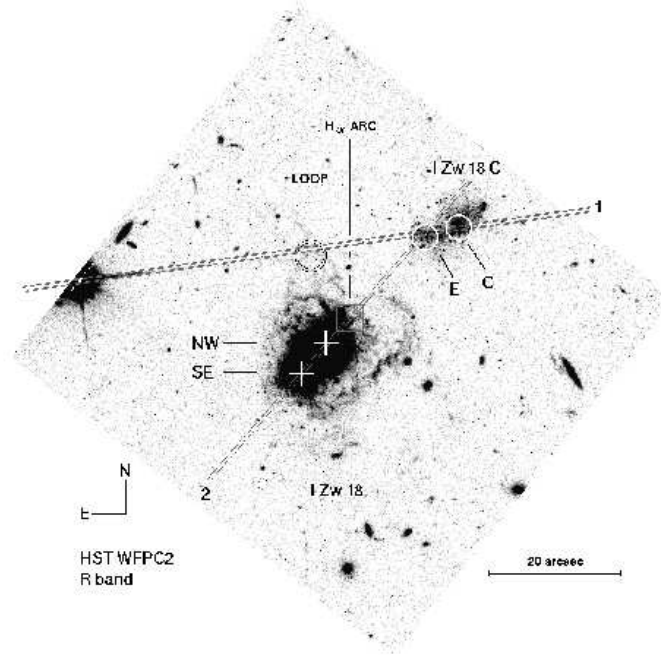}
\caption{\label{Fig1}
{\sl HST} archival image of I Zw 18 in the $R$ band. North is up, east 
to the left. The orientation of the slit
during Keck II observations is shown as ``1'', while the
orientation of the slit during the 4m KPNO and MMT spectroscopic 
observations is shown as ``2''. The two regions of star formation in I Zw 18
are marked as ``SE'' and ``NW''. The locations of the outer regions with strong 
gaseous emission in I Zw 18 are labeled as ``H$\alpha$ ARC'' and ``LOOP''.
The locations of the central and eastern regions of I Zw 18C are labeled 
respectively as ``C'' and ``E''.}
\end{figure}

\begin{figure}
\epsscale{0.8}
\figurenum{2}
\plotone{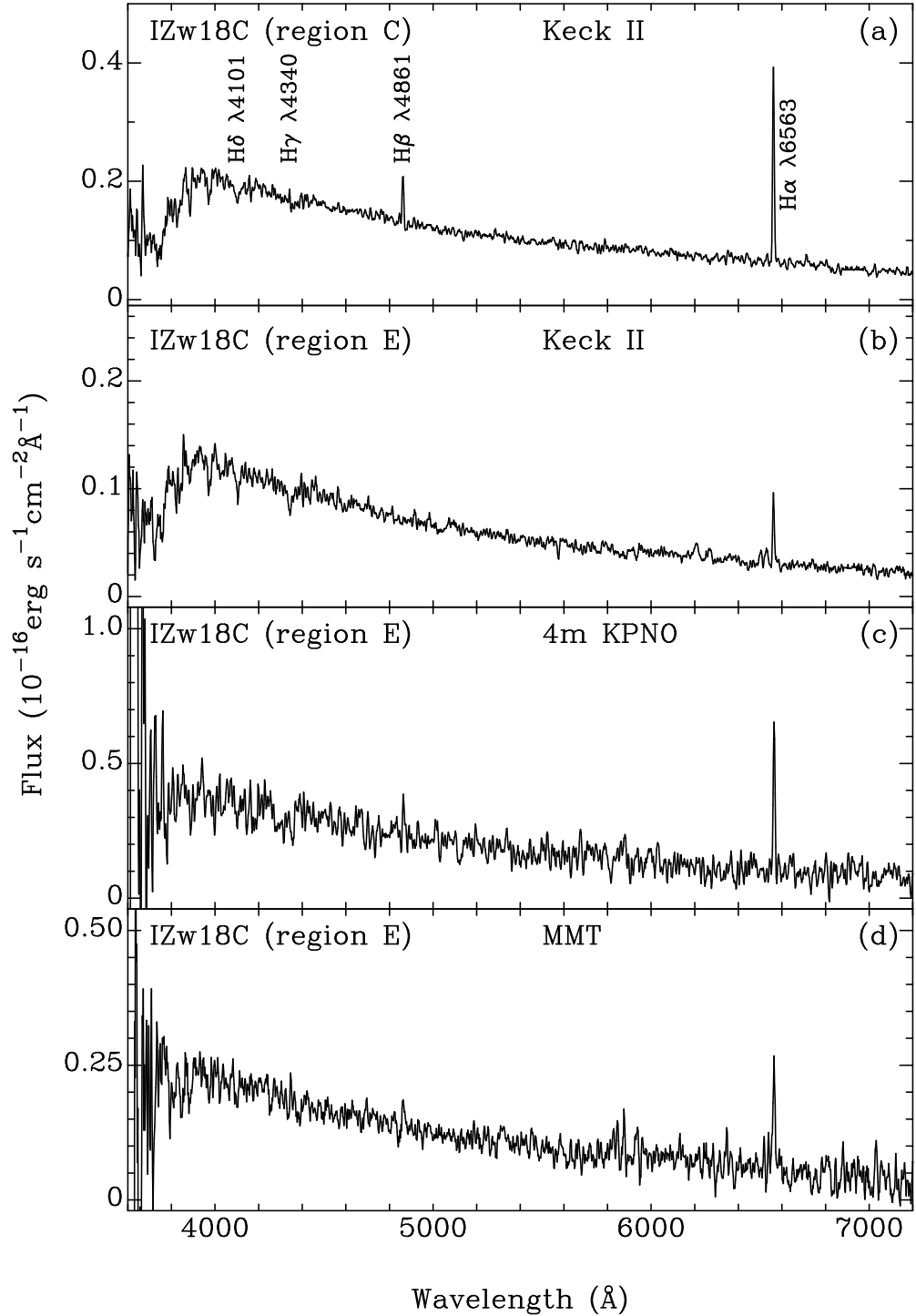}
\caption{\label{Fig2}
Keck II, 4m KPNO and MMT spectra of the central (C) 
and eastern (E) regions of I Zw 18C. Keck II spectra are extracted 
within a 0\farcs9$\times$1\arcsec\ aperture, the 4m KPNO and 
MMT spectra are extracted 
within 2\arcsec$\times$5\arcsec\ and 1\farcs5$\times$4\arcsec\ apertures
respectively. The hydrogen emission and
absorption lines are marked in the upper panel (a). All spectra are smoothed
by a 3-point box-car.}
\end{figure}

\begin{figure}
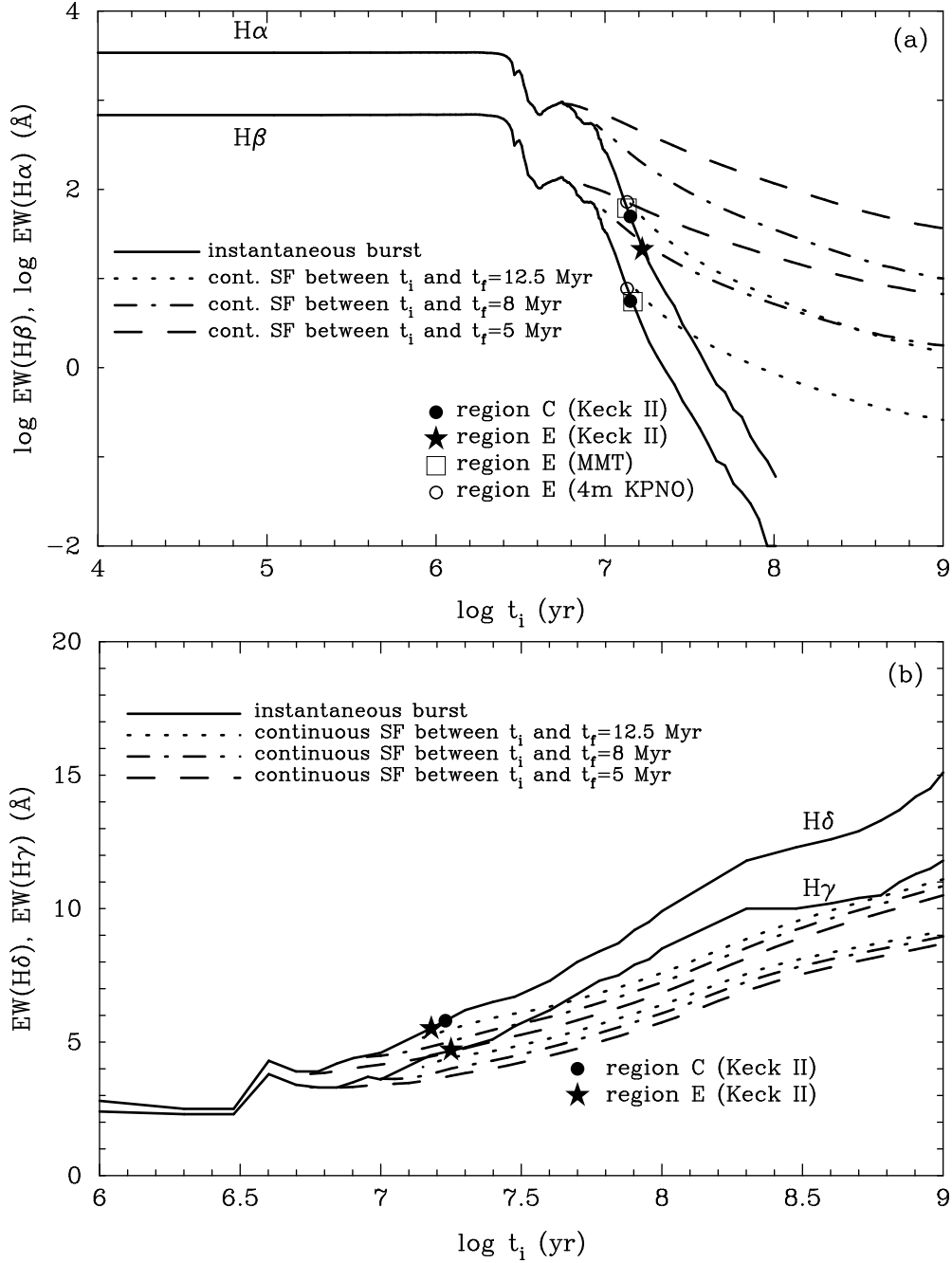

\epsscale{0.8}
\figurenum{3}
\plotone{Fig3a.ps}
\plotone{Fig3b.ps}
\caption{\label{Fig3}
(a) Model dependences of the H$\alpha$ and H$\beta$ emission line 
equivalent widths on age for an instantaneous burst with the heavy element 
mass fraction $Z$ = $Z_\odot$/50 (solid lines, Schaerer, private communication;
Lejeune \& Schaerer 2001). 
The measured H$\alpha$ and H$\beta$ equivalent widths are
shown by different symbols and are consistent with a single stellar population
age of $t$ = 15 Myr. Model predictions are also shown for constant continuous 
star formation starting at an initial time defined by 
the abscissa $t_{\rm i}$ and stopping at a final time $t_{\rm f}$, 
with $t_{\rm f}$ = 5 Myr (dashed line), 
$t_{\rm f}$ = 8 Myr (dot-dashed line) and $t_{\rm f}$ = 12.5 Myr (dotted line).
Time $t_{\rm i}$ = 0 is now and increases to the past.
(b) Model dependences of the H$\gamma$ (dashed line) and H$\delta$ 
(solid line) absorption line equivalent 
widths on age for an instantaneous burst with the heavy element mass
fraction $Z$ = $Z_\odot$/20 (Gonzalez Delgado et al 1999). 
The lines and symbols have the same meaning as in (a).}
\end{figure}

\begin{figure}
\epsscale{0.8}
\figurenum{4}
\plotone{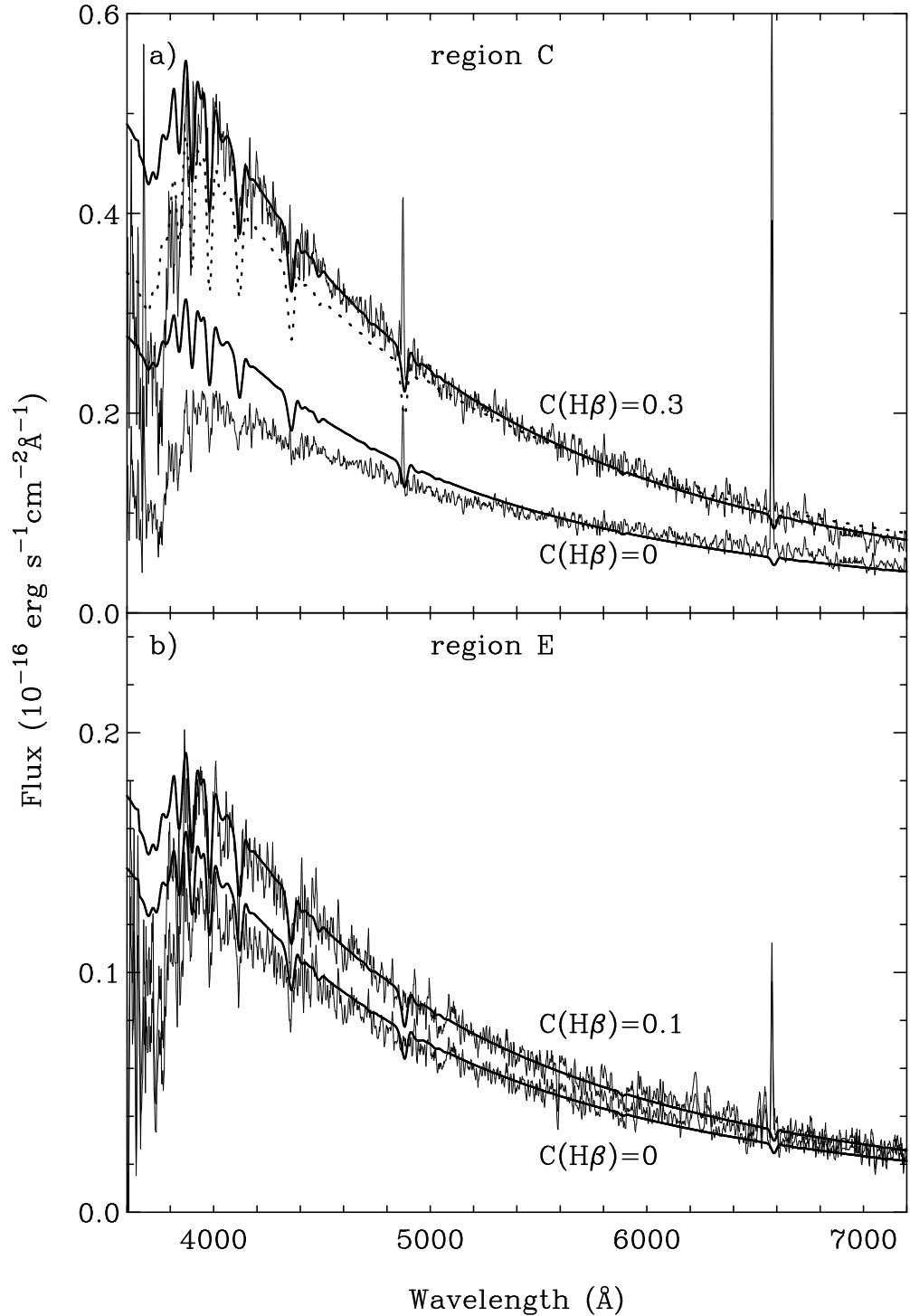}
\caption{\label{Fig4}
(a) Keck II spectra of region C in I Zw 18C corrected
for reddening with two values of the extinction coefficient 
$C$(H$\beta$) = 0 and 0.3 (thin solid lines) on which are
superposed model stellar population SEDs with age 15 Myr (thick solid lines).
The dotted line is the model stellar population SED with age 40 Myr.
(b) Keck II spectra of region E in I Zw 18C corrected
for the reddening with two values of the extinction coefficient 
$C$(H$\beta$) = 0 and 0.1 (thin solid lines) on 
which are superposed model stellar population SEDs with age 15 Myr 
(thick solid line).}
\end{figure}

\begin{figure}
\epsscale{0.9}
\figurenum{5}
\plotone{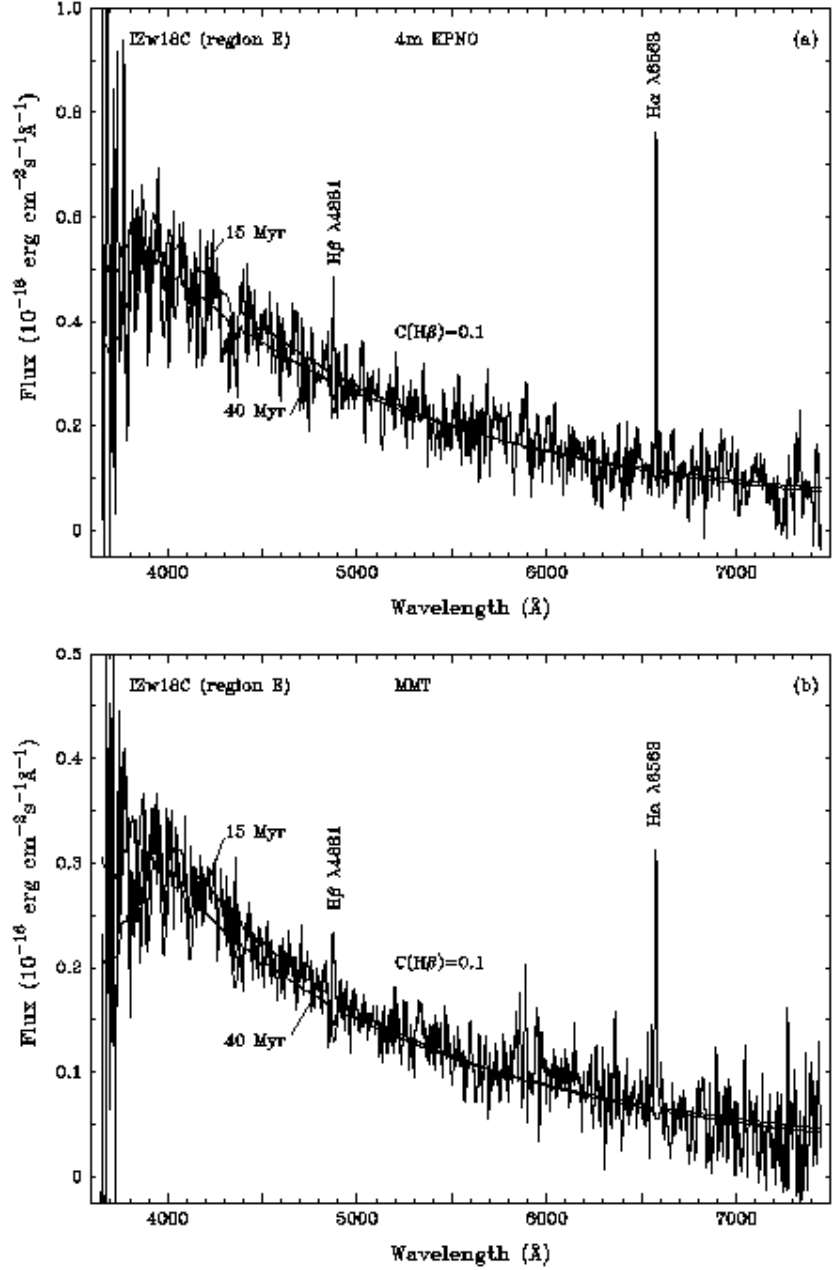}
\caption{\label{Fig5}
(a) 4m KPNO and (b) MMT spectra of region E in I Zw 18C (thin line) on which are
superposed model stellar population SEDs with ages 15 Myr and 40 Myr (thick 
lines). The observed spectra are corrected for extinction with 
$C$(H$\beta$) = 0.1.}
\end{figure}

\begin{figure}
\epsscale{0.6}
\figurenum{6}
\plotone{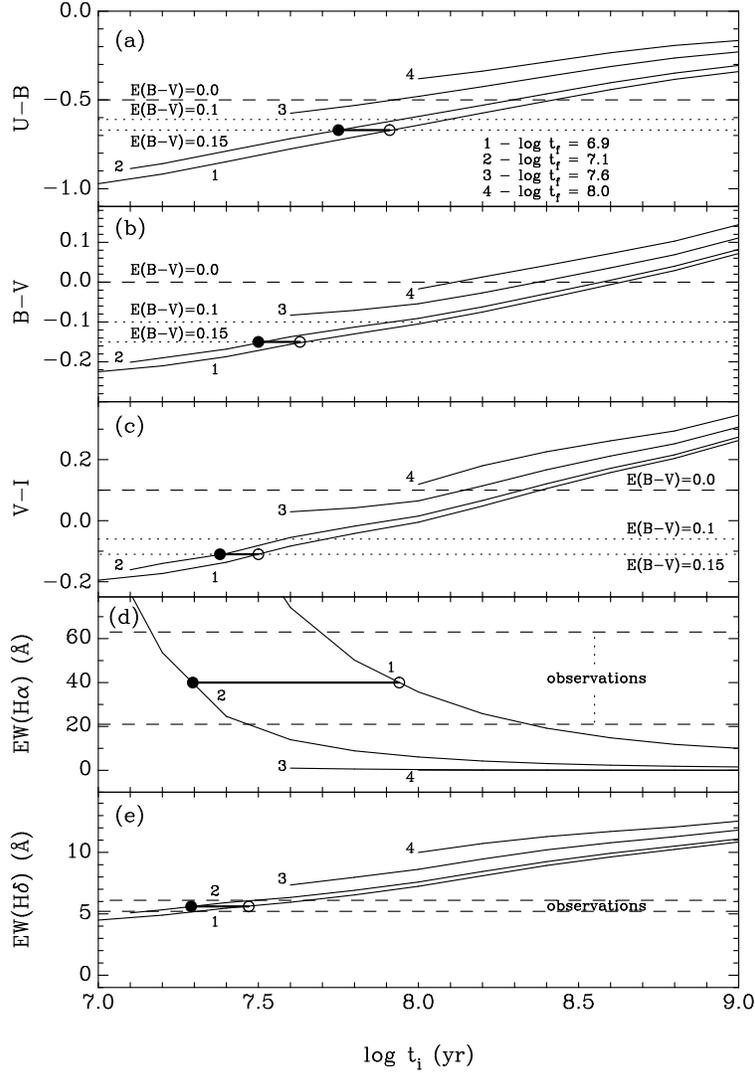}
\caption{\label{Fig6}
(a) - (c) Model dependences of the $(U-B)$, $(B-V)$, $(V-I)$ colors on
age, for constant continuous star formation starting at an initial time
defined by the abscissa $t_{\rm i}$ and stopping at a final time $t_{\rm f}$,
with a heavy element mass fraction $Z$ = $Z_\odot$/50 (solid lines, Schaerer,
private communication; Lejeune \& Schaerer 2001). Curves with
different $t_{\rm f}$ given in panel (a) 
are labeled from 1 to 4. The dashed lines
show observed colors without correction for interstellar extinction. The
dotted lines show colors corrected for interstellar extinction with 
$E$($B-V$) = 0.1 and 0.15. 
Open and filled circles mark the observed colors  
corrected for extinction with $E$($B-V$) = 0.15 on the model lines   
with star formation stopping at log $t_{\rm f}$ =
6.9 and 7.1, where $t_{\rm f}$ is in yr. 
(d) - (e) Model dependences of
the H$\alpha$ emission line ($Z$ = $Z_\odot$/50, Schaerer,
private communication; Lejeune \& Schaerer 2001) and H$\delta$ absorption line 
($Z$ = $Z_\odot$/20, Gonzalez Delgado et al. 1999) equivalent widths
on age, for constant continuous star formation starting at an initial time
defined by the abscissa $t_{\rm i}$ and stopping at a final time $t_{\rm f}$. 
The dashed lines show the range of observed equivalent widths. Open and filled
circles show the location of the mean observed equivalent widths on
the model lines for log $t_{\rm f}$ = 6.9 and 7.1.}
\end{figure}

   Ground-based and {\sl HST} H$\alpha$ and broad-band imaging have revealed
filamentary structure around I Zw 18, inside a 15\arcsec\ radius 
(e.g., Hunter \& Thronson 1995; \"Ostlin, Bergvall \& R\"onnback 1996; 
Dufour et al. 1996b). Because of the presence of ionizing young stars, 
the light in these outlying regions is likely to be dominated by
the emission of the ionized gas. This conclusion is supported by 
spectroscopic observations. Dufour et al. (1996a) and Izotov \& Thuan
(1998) have detected H$\alpha$ emission at distances as large as 20\arcsec\ 
from I Zw 18 in the NW direction of slit ``2'' (Fig. \ref{Fig1}).
Izotov et al. (1999) have found that at
a distance of $\sim$ 5\arcsec\ to the northwest of the brightest NW region
of I Zw 18, the equivalent width of the H$\beta$ emission line is $\sim$ 300\AA.
In this case the contribution of the gaseous continuum near H$\beta$ is 
$\sim$ 30 percent of the total continuum. The contribution of the gaseous 
continuum near H$\alpha$ is even larger, being $\sim$ 50 percent
of the total continuum. Therefore, when analyzing stellar populations
with the use of photometric data, it is essential to correct broad-band colors 
for ionized gas emission. 

\begin{figure}
\epsscale{0.8}
\figurenum{7}
\plotone{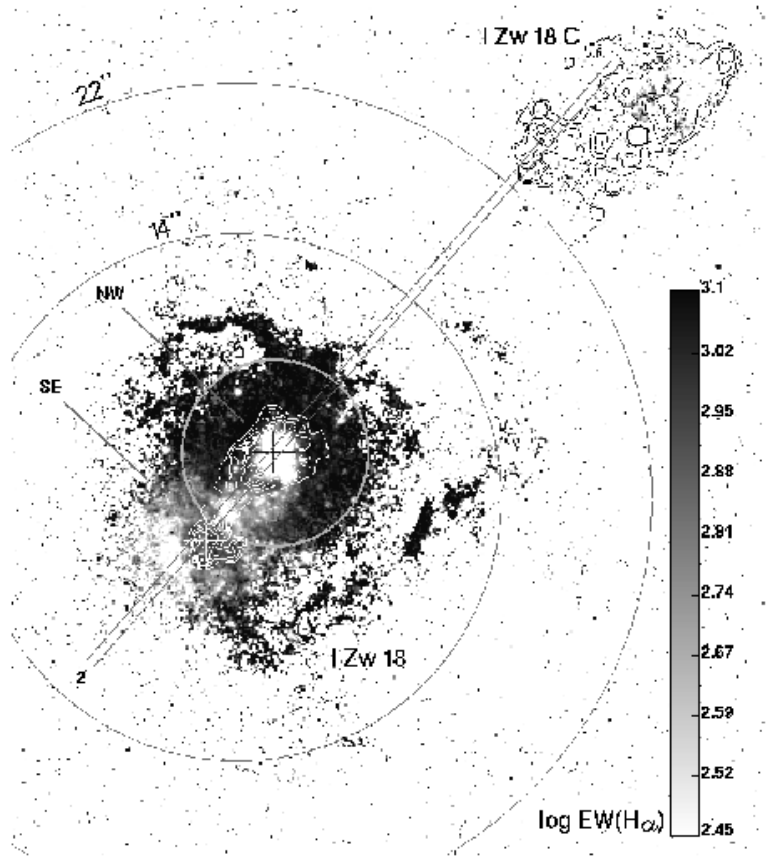}
\caption{\label{Fig7}
H$\alpha$ equivalent width map of I Zw 18 in the
range between 280 \AA\ and 1260 \AA. The H$\alpha$ equivalent width
increases from $\sim$100\AA\ at the location of the NW star-forming
region to $\sim$1300\AA\ roughly 5\arcsec\ to the northeastern
direction. This region, referred to as ``H$\alpha$ arc'' 
(see Fig. \ref{Fig1})
coincides roughly with the intersection of the overlaid thick-grey
circle with slit ``2''. The larger circles with radii
14\arcsec\ and 22\arcsec\ are centered between the SE and NW regions.}
\end{figure}

However, in some recent papers (e.g., Kunth \& \"Ostlin 2000) 
the extended emission around I Zw 18 has been attributed to an old stellar 
population, while the contribution of the ionized gas is assumed to be not dominant. 
That this cannot be true is seen in Fig. \ref{Fig7} where we show a 
map of the H$\alpha$ equivalent width distribution as obtained from {\sl HST} 
narrow-band and broad-band images. While the H$\alpha$ equivalent width is small 
in the direction of the stellar clusters, it exceeds 1000\AA\ in the outer
regions.

In the following, we analyze MMT and Keck II spectroscopic observations of 
the outer regions around I Zw 18 to clarify two issues: a) how 
important is the contribution of the ionized gas in the outer regions of
I Zw 18?  and b) is stellar emission present at large distances?

\begin{figure}
\epsscale{0.9}
\figurenum{8}
\plotone{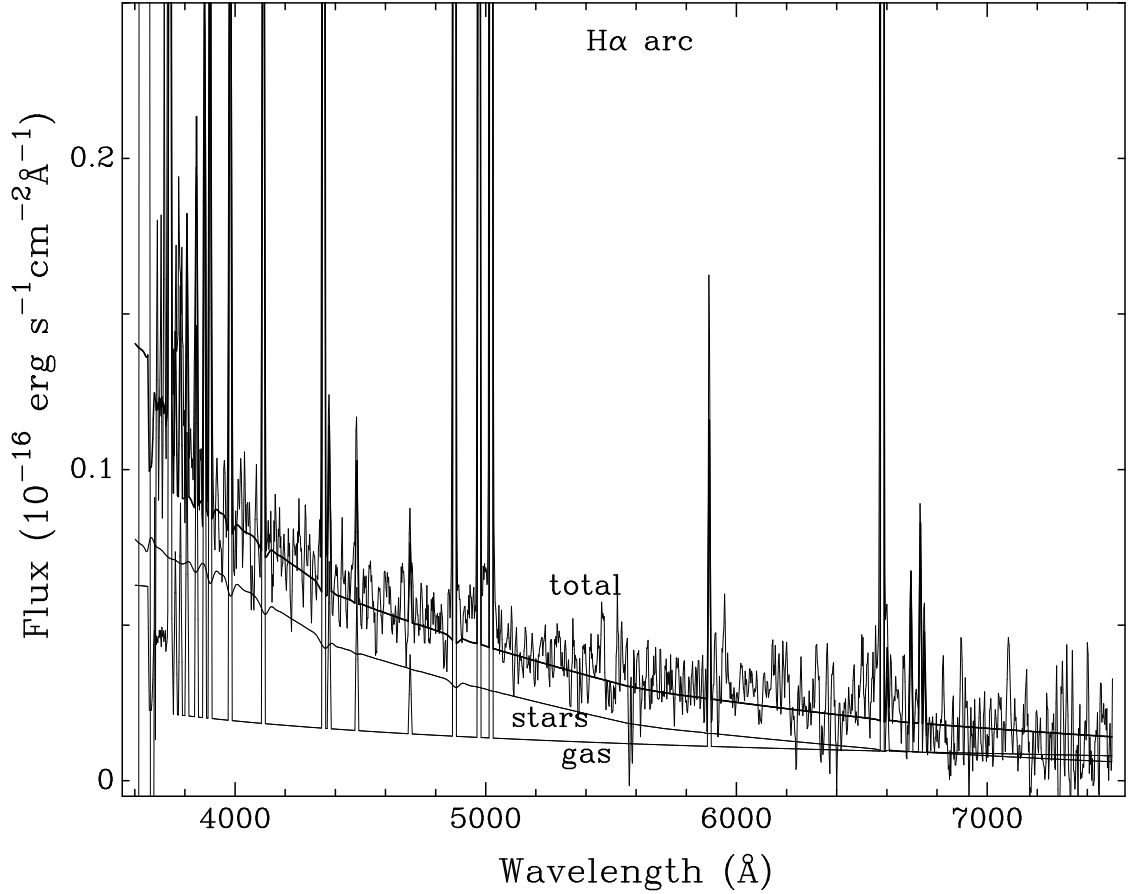}
\caption{\label{Fig8}
MMT spectrum of the H$\alpha$ arc in I Zw 18 (thin line) on which is 
superposed a synthetic continuum including both gaseous and stellar emission 
from a starburst with age 2 Myr (thick line). The spectrum is extracted 
within a 1\farcs5 $\times$ 3\arcsec\ aperture, centered at a distance 
5\arcsec\ to the northwest from the NW component of I Zw 18 (in slit
``2''). It is 
corrected for an extinction $A_V$ = 0.13 mag, derived from the Balmer 
hydrogen emission line decrement.}
\end{figure}

In Figure \ref{Fig8} we show the MMT spectrum of the region with a high H$\beta$
equivalent width. The spectrum is extracted 
within an aperture 1\farcs5 $\times$ 3\arcsec\ (slit ``2''), centered 
at a distance 
5\arcsec\ to the northwest from the NW component of I Zw 18. It is
characterized by strong emission lines. We refer to this region as 
the ``H$\alpha$ arc'' (square box in Fig. \ref{Fig1}).
A synthetic spectrum with a 2 Myr stellar population combined with ionized gas 
emission fits best the observed SED of this region.
The observed and extinction-corrected 
emission-line intensities in the H$\alpha$ arc together with their equivalent
widths are listed in Table \ref{Tab4}.

The ionic and elemental abundances have been derived following
Izotov et al. (1994, 1997c). The extinction coefficient 
$C$(H$\beta$) and the absorption equivalent width $EW$(abs) for the hydrogen 
lines are obtained by an iterative procedure. They are shown in Table 
\ref{Tab4} together with the observed flux $F({\rm H}\beta)$
of the H$\beta$ emission line.
The electron temperature $T_e$(O {\sc iii}) was determined
from the [O {\sc iii}] $\lambda$4363 / ($\lambda$4959 + $\lambda$5007) flux
ratio and the electron number density $N_e$(S {\sc ii}) from the 
[S {\sc ii}] $\lambda$6717/$\lambda$6731 flux ratio.  The ionic and elemental
abundances are shown in Table \ref{Tab5} together with ionization correction 
factors (ICFs). They are in good agreement with the abundances derived by 
Skillman \& Kennicutt (1993), Izotov \& Thuan (1998), V\'ilchez \& 
Iglesias-P\'aramo (1998) and Izotov et al. (1999) for the NW and SE components
of I Zw 18.

\begin{figure}
\epsscale{0.8}
\figurenum{9}
\plotone{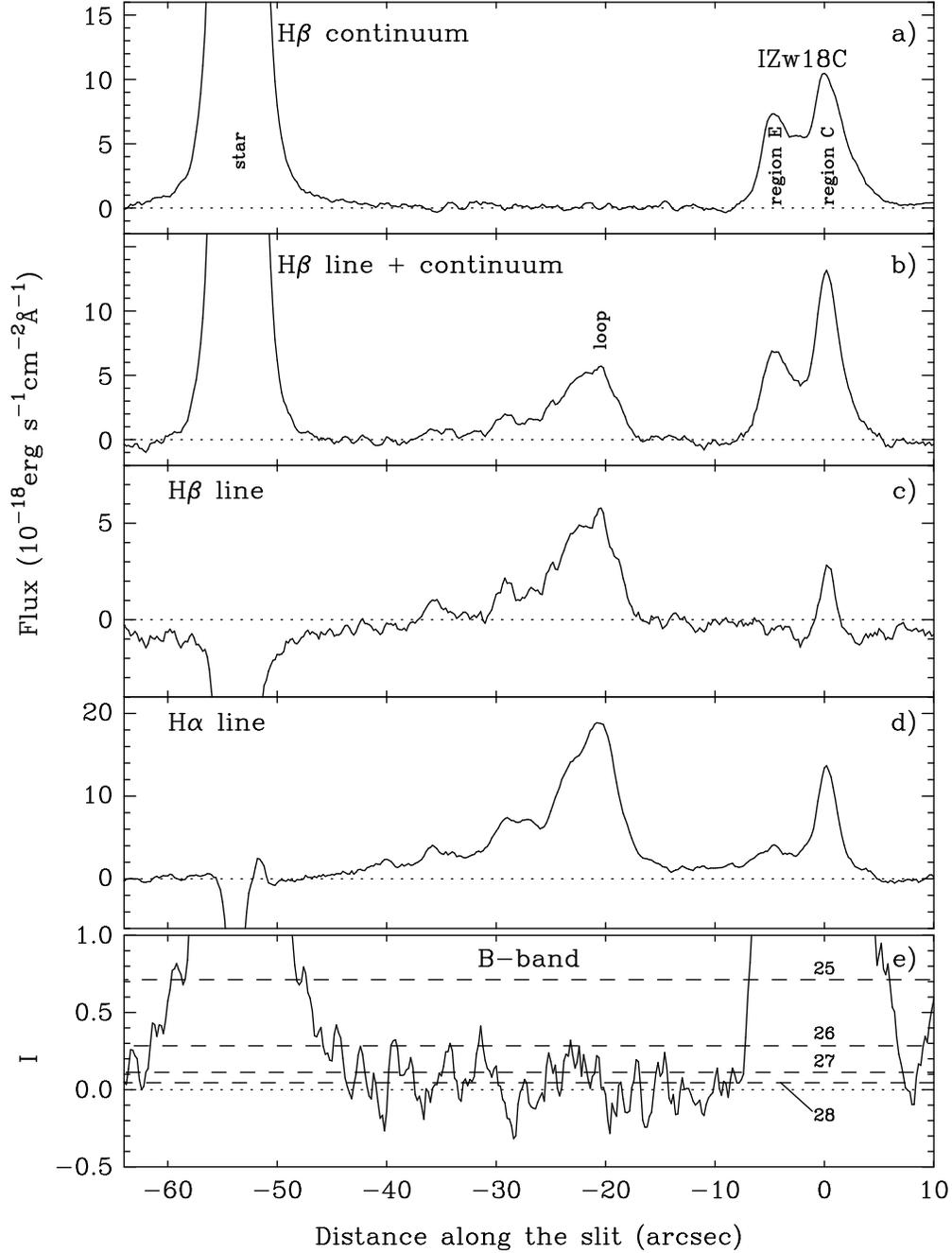}
\caption{\label{Fig9}
(a) Distribution of the H$\beta$ continuum flux along the slit
from Keck II spectra
with position angle --80$^\circ$ (slit ``1'' in Fig. \ref{Fig1}). 
The locations of the E and C regions in I Zw 18C and the
bright star (at the edge of Fig. \ref{Fig1}) are marked.
(b) The distribution of the H$\beta$ line + continuum flux along the 
slit. The location of the expanding shell is marked as ``loop''.
(c) Distribution of the continuum-subtracted H$\beta$ line intensity 
along the slit.
(d) The distribution of the continuum-subtracted H$\alpha$ line flux 
along the slit. Note that H$\alpha$ emission is detected at a distance
$\ga$ 40\arcsec\ from region C of IZw18C, or at a distance as large as $\sim$
30\arcsec\ from I Zw 18.
(e) Distribution along the slit of the continuum intensity at 
$\lambda$4200 in units 10$^{-18}$ erg s$^{-1}$cm$^{-2}$\AA$^{-1}$arcsec$^{-2}$. 
Different levels of the $B$ surface brightness in mag 
arcsec$^{-2}$ are shown by dashed lines.}
\end{figure}

We have shown that the contribution of the ionized gas is large in the 
H$\alpha$ arc, at $\sim$ 5\arcsec\ from the NW component of I Zw 18.
A similar situation prevails at significantly larger distances, as evidenced by
deep Keck II spectroscopic observations. The slit during these observations 
(slit ``1'' in  Fig. \ref{Fig1}) crossed the outer regions of I Zw 18 
including the expanding supershell of ionized gas best seen in H$\alpha$ images
(e.g., Hunter \& Thronson 1995; Dufour et al. 1996b). The latter
feature located at $\sim$ 15\arcsec\ from I Zw 18 is labeled in Figure 
\ref{Fig1} as ``LOOP''. We can thus study 
with deep spectroscopy the extended diffuse emission in I Zw 18 all 
the way from I Zw 18C to the bright star at the edge of Fig. \ref{Fig1} and 
located $\sim$ 40\arcsec\ from I Zw 18. In Fig. \ref{Fig9}a -- \ref{Fig9}b
we show respectively the flux distributions along the slit of the continuum
near H$\beta$ and of the line + continuum emission at the wavelength
of the H$\beta$ emission line. The origin is taken to be at region C. 
The locations of the bright star and loop are marked. In Fig. \ref{Fig9}c we 
show the continuum-subtracted flux distribution of the H$\beta$ emission line.
The negative values of the flux in some regions of I Zw 18C and
around the star are probably caused by underlying stellar H$\beta$ 
absorption. No appreciable continuum
emission is seen in Fig. \ref{Fig9}a between I Zw 18C and the star.
However, nebular emission of H$\alpha$ is present nearly
everywhere between I Zw 18C and the star (Fig. \ref{Fig9}c -- \ref{Fig9}d), 
suggesting that the
contribution to the total flux of the nebular emission from ionized gas
is important in the outermost regions, as far as 30\arcsec\ from I Zw 18. 

In Fig. \ref{Fig9}e we show the intensity distribution of the continuum at 
the wavelength of 4200\AA\ approximating the $B$ band. We also plot by 
dotted lines the surface brightness levels in steps of 1 mag 
arcsec$^{-2}$. It is seen that the continuum surface brightness is fainter 
than 27 mag arcsec$^{-2}$ everywhere between I Zw 18C and the star, including 
the loop region. However, we cannot 
exclude the presence of stellar emission at the level of 28 mag arcsec$^{-2}$, 
as postulated by Legrand (2000). The contamination by extended ionized gas 
emission makes the detection of such an extremely faint hypothetical
stellar background problematic.

\begin{figure}
\epsscale{0.9}
\figurenum{10}
\plotone{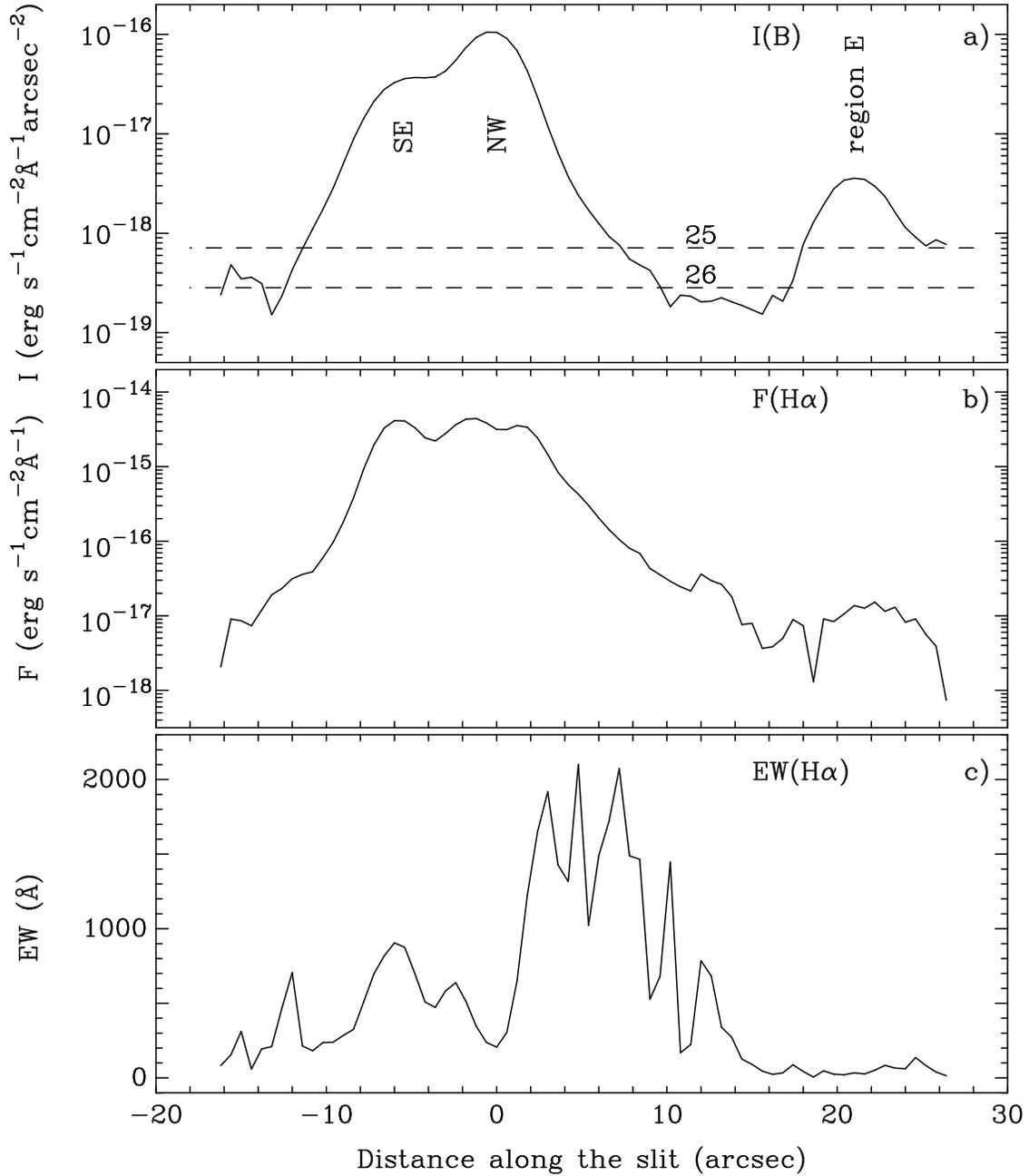}
\caption{\label{Fig10}
(a) Distribution of the continuum intensity at 
$\lambda$4200 along the slit from MMT spectra with position angle --41$^\circ$ 
(slit ``2'' in Fig. \ref{Fig1}). Different levels of the $B$ surface brightness 
in mag arcsec$^{-2}$ are shown by dashed lines. 
The distribution is smoothed by a 5-point box-car.
The locations of the SE, NW components of I Zw 18 and region E of I Zw 18C
are marked.
(b) Distribution of the continuum-subtracted H$\alpha$ line flux 
along the slit. Note that the H$\alpha$ emission in the main body is more 
extended as compared to the continuum distribution.
(c) Distribution of the H$\alpha$ equivalent width. Note the large values
of $EW$(H$\alpha$) to the northwest of the NW component implying an important
contribution of gaseous emission to the total light in the extended region
around the main body.}
\end{figure}

In Fig. \ref{Fig10} we show the distributions along slit ``2'' 
of: a) the continuum intensity at 4200\AA, b) the
continuum-subtracted flux and c) the equivalent width of the H$\alpha$ 
emission line. The distribution of H$\alpha$ emission (Fig. \ref{Fig10}b) around
the main body is more extended as compared to the continuum (Fig. \ref{Fig10}a),
the latter being confined in a region with radius less than 12\arcsec\ 
around the NW component. The equivalent width of H$\alpha$ is very high to
the northwest of the NW component (Fig. \ref{Fig10}c) and must be taken
into account when photometric properties of the stellar population 
in I Zw 18 are analyzed. We also point out that the continuum distribution
of region E in I Zw 18C is narrower than that of the H$\alpha$ emission line.
The maximum H$\alpha$ equivalent width in region E 
(Fig. \ref{Fig10}c) is offset to the northwest 
by $\sim$ 2\arcsec\ relative to the continuum distribution (Fig. \ref{Fig10}a).

\begin{figure}
\epsscale{0.9}
\figurenum{11}
\plotone{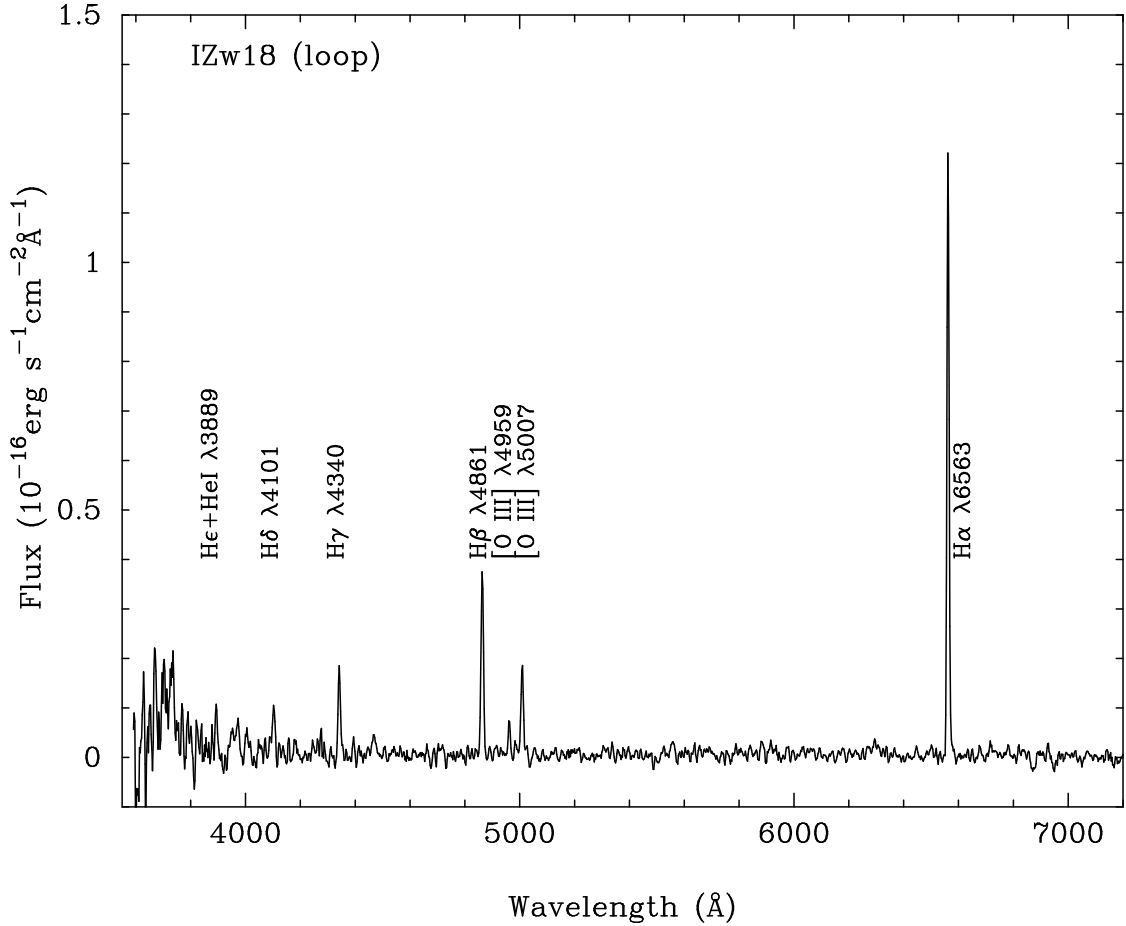}
\caption{\label{Fig11}
Spectrum of the expanding shell of I Zw 18  obtained 
with the Keck II (slit ``1'' in Fig. \ref{Fig1}) 
and extracted in a 0\farcs9$\times$10\arcsec\ aperture at the distance of 
$\sim$ 15\arcsec\ from I Zw 18. The spectrum is smoothed by a 3-point box-car.}
\end{figure}

In Fig. \ref{Fig11} we show the spectrum of the loop. Despite its faintness,
several emission lines are seen. However, the sensitivity in the blue region was 
not sufficient to detect the [O {\sc ii}] $\lambda$3727 emission line.
The continuum is very weak and can be significantly affected by uncertainties in 
the sky subtraction. This makes the measurements of line equivalent widths 
difficult. The fluxes and equivalent widths of the detected lines are given in 
Table \ref{Tab6}. The flux errors include uncertainties in the placement of 
the continuum level and in the fitting of the lines by Gaussian profiles. 
However, these errors do not take into account the uncertainties in the sky 
subtraction which might be large. Indeed, the loop flux in the continuum 
is only $\sim$ 1\% above the night sky flux, while that number is as high as
50\% for the continuum flux in I Zw 18C. Even with these large uncertainties 
the emission line equivalent widths in the loop spectrum
are very high. In particular, the 
equivalent width of the H$\beta$ emission line is 471 \AA\ or about half of 
the value expected for pure gaseous emission at the electron temperature
$T_e$ = 20000K. Hence, half of the flux in the continuum comes from the
ionized gas, emphasizing again the importance of the 
correction of the spectral energy
distribution and broad-band colors for gaseous emission. This goes contrary to
the assumption of Kunth \& \"Ostlin (2000) that the contribution of gaseous
emission does not affect the colors of the outlying regions of 
I Zw 18. If errors in the night sky subtraction are $\sim$ 1\%, then 
the equivalent width of the H$\beta$ emission line in the loop spectrum
is in the range $\sim$ 250 -- 1000\AA. Within the uncertainties, the emission 
of the loop is quite consistent with pure gaseous emission.

Kunth \& \"Ostlin (2000) have derived radial distributions of the 
surface brightness in the $B$ band and of the $B-R$ and $B-J$ colors of I Zw 18 
(their Fig. 8). They find that the colors rise continuously with 
increasing radius and reach $B-R$ = 0.6 mag and $B-J$ = 1.6 mag at a radius of
10\arcsec. Assuming a purely stellar emission, they conclude that the
observed colors can be reproduced by a single stellar population model with a 
metallicity of 1/50 $Z_\odot$ and 
an age of log $t$ = 9.1 $\pm$ 0.1 ($t$ in yr), irrespective of the IMF
(Bruzual \& Charlot 2000, unpublished). 
However, this age estimate is rather uncertain and is dependent on the
particular population synthesis model used. For example, Tantalo et al. (1996) 
using Padua stellar evolutionary tracks give values $B-R$ = 0.8 mag
and $B-J$ = 1.7 mag for a 1 Gyr single stellar population
with a metallicity of 1/50 $Z_\odot$. The bluer colors derived by Kunth 
\& \"Ostlin (2000) would give an age of 100 -- 300 Myr according to 
Tantalo et al. (1996)' models. On the other hand, Leitherer et al. (1999)'
models using the Geneva stellar evolutionary tracks predict $B-R$ = 0.5 mag and 
$B-J$ = 1.1 mag for a 1 Gyr single stellar 
population with a metallicity of 1/20 $Z_\odot$, bluer than those
derived by Kunth \& \"Ostlin (2000). Leitherer et al. (1999)' models do not
go beyond 1 Gyr, but to reproduce the colors derived by 
Kunth \& \"Ostlin (2000), the age of the stellar population in the outer regions 
of I Zw 18, if present, must be older than 1 Gyr. Their models are calculated
for a metallicity of 1/20 $Z_\odot$, but colors with a metallicity of 
1/50 $Z_\odot$ are expected to be bluer, further increasing the derived age.
These age estimates are very uncertain. The models by Leitherer et al. 
(1999) are less reliable for ages greater 100 Myr because they do not include 
asymptotic giant branch (AGB) star evolution. Tantalo et al. (1996) do include
AGB star evolution, but the little known mass loss processes in the 
AGB phase introduce uncertainties in the predicted colors (Girardi \& Bertelli 
1998).

The next source of uncertainties comes from the photometric observations
themselves. Beyond a radius of $\sim$ 5\arcsec\ from I Zw 18, the $B-R$ color 
profile derived by Kunth \& \"Ostlin (2000) increases monotonously while the 
$B-J$ color profile shows discontinuous jumps. These discontinuities are 
difficult to understand if the same stellar population is responsible for both
colors. Kunth \& \"Ostlin (2000) do not show the
uncertainties of their photometry. However, similar deep $J$-band photometry
of another galaxy SBS 0335--052 (Vanzi et al. 2000) with UKIRT shows that
at the $J$-band surface brightness of 24 -- 25 mag arcsec$^{-2}$ the errors
are already $\sim$ 0.5 mag or more. New recent $B$ and $J$ photometric 
observations of I Zw 18 (Papaderos et al. 2001) do not confirm the large 
reddening of the $B-J$ color observed by Kunth \& \"Ostlin (2000) between radii 
6\arcsec\ and 8\arcsec, nor the discontinuous jumps.

To investigate whether the $B-R$ and $B-J$ colors of the extended emission 
can be explained by pure gaseous emission, we calculate the spectral energy 
distribution of the
ionized gas emission in the corresponding wavelength range. 
The contribution of the
free-bound, free-free and two-photon continuum emission is taken into
account for the spectral range from 0 to 5 $\mu$m (Aller 1984; 
Ferland 1980). As for the electron temperature, we adopt the value of
19000K, which is the mean value between the electron temperatures in the
NW and SE components of I Zw 18. Emission lines are superposed on the 
gaseous continuum SED with intensities derived from the observed spectrum 
of the loop at the distance of $\sim$ 15\arcsec\ from I Zw 18 (Table
\ref{Tab6}), in the spectral range 
$\lambda$3700 -- 7500 \AA. Outside this range, the intensities of emission 
lines (mainly hydrogen lines) have been calculated from the 
extinction-corrected flux of H$\beta$ with reddening $A_V$ = 0.16 mag. 
The reddening in the loop was calculated from the observed H$\alpha$/H$\beta$
flux ratio (Table \ref{Tab6}), assuming an electron temperature $T_e$ = 20000K.
We derive $B-R$ = 0.8 mag and $B-J$ = 0.9 mag. If instead of the relative 
intensities of the emission lines observed in the loop, we use those seen
in the NW or SE regions of I Zw 18 (Izotov et al. 1999), we obtain slightly 
bluer colors, $B-R$ = 0.6 mag and $B-J$ = 0.7 mag. The color difference
is mainly due to a smaller contribution in the outer regions 
of some emission lines, e.g. [Ne {\sc iii}] $\lambda$3869 to the $B$ band. 
From this comparison we conclude that colors become redder at larger distances,
even in the case of pure gaseous emission.
While the $B-R$ color of gaseous emission
is similar to the asymptotic value of $\sim$ 0.7 mag derived by
Kunth \& \"Ostlin (2000) at distances $\sim$ 15\arcsec, the 
predicted $B-J$ color of gaseous emission is considerably bluer than the value
they obtained. However, that value is consistent with $B-J$ $\sim$ 0.6 mag
derived  by Papaderos et al. (2001) in the 6 -- 9 arcsec radius range.

We note that the $B-R$ and $B-J$ colors are not ideal for constraining the
existence of a possible extended low-surface-brightness 1 Gyr 
underlying stellar population in I Zw 18, the 
latter being uncertain, and the former being very similar to the color of
ionized gas. The $B-I$ color is more useful because it can discriminate
better between gaseous and a 1 Gyr stellar population
emission. Indeed, adopting the
relative line intensities in the H$\alpha$ arc or in the loop,
the $B-I$ colors of the ionized gas emission are --0.1 mag and +0.1 mag
respectively. The expected $B-I$ color for a 1 Gyr stellar population is
much redder, $\sim$ +1.2 mag (Tantalo et al. 1996). Observations give
$B-I$ $\sim$ 0 at radii 8 -- 10 arcsec (Papaderos et al. 2001), strongly 
suggesting that the emission in the outer parts of I Zw 18 is gaseous in origin.

We conclude that there is no convincing observational evidence for
the presence of an extended underlying low-surface-brightness stellar
component in I Zw 18. Its existence, as postulated by Kunth \& \"Ostlin (2000),
Legrand (2000) and Legrand et al. (2000), is neither supported by spectroscopic 
nor photometric observations.

\section{Conclusions} 

We use spectroscopic and photometric data to constrain 
the age of the stellar population in 
the C component of I Zw 18 ($\equiv$ I Zw 18C) and to study the 
origin of the extended emission around I Zw 18.
We have arrived at the following main conclusions:

1. Deep 4m KPNO, MMT and Keck II spectra of I Zw 18C 
show H$\beta$ and H$\alpha$ hydrogen lines in 
emission, and H$\delta$ and H$\gamma$ hydrogen lines in 
absorption. Using their equivalent widths we derive an age of the stellar 
population of $\sim$ 15 Myr if an instantaneous burst is assumed.
If star formation is continuous, then the equivalent widths are best reproduced 
by a short star formation episode continuously occurring between $\sim$ 10 Myr and 
$\sim$ 25 Myr ago. Uncertainties in the observations and models may extend
this range to between $\sim$ 10 Myr and $\la$ 100 Myr ago.

2. Spectral energy distributions of the central (C) and eastern (E) regions of 
I Zw 18C are used to derive extinction.
The equivalent widths of the hydrogen emission 
and absorption lines and the spectral energy distributions are
modeled by a 15 Myr single stellar population if the extinction
coefficient $C$(H$\beta$) = 0.1 -- 0.3, corresponding to
$A_V$ = 0.20 -- 0.65 mag.

3. With the usually assumed distance of $\sim$ 10 Mpc the stellar population
age derived from {\sl HST} color-magnitude diagrams is too large as compared to 
the young age derived from the spectroscopic data.
One possible source of the difference is interstellar
extinction. Furthermore, to have agreement between the CMDs and the ionization 
state of I Zw 18C, the distance to the BCD should be increased to $\sim$ 15 Mpc.

4. Concerning the extended emission around I Zw 18, Keck II spectra show
H$\alpha$ emission as far as 30\arcsec\ from the main body. 
The equivalent widths of emission lines are particularly strong in the extended
envelope ($EW$(H$\beta$) = 471\AA), implying a dominant contribution of 
the ionized gas emission in the outermost regions of I Zw 18. Within the 
large uncertainties of the continuum level, the emission at 
$\sim$ 15\arcsec\ from I Zw 18 is consistent with pure ionized gas emission.
We do not find evidence for an old extended 
low-surface-brightness stellar component in the outlying regions of I Zw 18
down to the surface brightness level $B$ $\sim$ 27 mag arcsec$^{-2}$, 
contrary to suggestions by Kunth \& \"Ostlin (2000).
It will be very difficult to detect the extended stellar emission at
$B$ $\sim$ 28 mag arcsec$^{-2}$ postulated by Legrand (2000), because of the
important ionized gas emission at large distances.

\acknowledgements 
Y.I.I. and N.G.G. thank the Universit\"ats--Sternwarte of G\"ottingen 
for warm hospitality.
We are grateful to D. Schaerer for making available his stellar 
evolutionary synthesis models in electronic form and for valuable comments
on the manuscript.
Y.I.I. thanks the G\"ottingen Academy of Sciences for a Gauss professorship.
We acknowledge the financial support of the Volkswagen Foundation Grant 
No. I/72919
(Y.I.I., N.G.G., P.P. and K.J.F.), of DFG grant 436 UKR 17/1/00 (N.G.G.),
Deutsche Agentur 
f\"{u}r Raumfahrtangelegenheiten (DARA) GmbH grants 50\ OR\ 9407\ 6
and 50\ OR\ 9907\ 7 (K.J.F. and P.P.), and of the National Science Foundation grants
AST-9616863 (T.X.T. and Y.I.I.) and AST-9803072 (C.B.F.).

\clearpage

\begin{deluxetable}{lcrccrccrccr}
\tablenum{1}
\tiny
\tablecolumns{12}
\tablewidth{0pt}
\tablecaption{Parameters of emission lines in I Zw 18C \label{Tab1}}
\tablehead{
    &\multicolumn{5}{c}{Keck  II}&&\multicolumn{2}{c}{MMT}
    &&\multicolumn{2}{c}{4m KPNO} \\ \cline{2-6} 
    \cline{8-9} \cline{11-12}
    &\multicolumn{2}{c}{region C}&&\multicolumn{2}{c}{region E}&&\multicolumn{2}{c}{region E}
    &&\multicolumn{2}{c}{region E} \\ \cline{2-3} \cline{5-6} \cline{8-9} 
    \cline{11-12}
Line&$F$\tablenotemark{a}&$EW$\tablenotemark{b}
&&$F$\tablenotemark{a}&$EW$\tablenotemark{b}
&&$F$\tablenotemark{a}&$EW$\tablenotemark{b}
&&$F$\tablenotemark{a}&$EW$\tablenotemark{b} }
\startdata
H$\beta$ &0.69$\pm$0.05& 5.6$\pm$0.3&& \nodata &  \nodata    &&
         0.76$\pm$0.28& 5.6$\pm$2.1&&1.50$\pm$0.28& 6.7$\pm$2.1 \\
H$\alpha$&3.08$\pm$0.06&49.7$\pm$1.0&&0.61$\pm$0.03&21.1$\pm$1.1&&
         2.78$\pm$0.26&51.8$\pm$4.9&&5.72$\pm$0.26&62.5$\pm$4.9 \\
\enddata
\tablenotetext{a}{in units of 10$^{-16}$ erg s$^{-1}$cm$^{-2}$.}
\tablenotetext{b}{in \AA.}
\end{deluxetable}

\clearpage

\begin{deluxetable}{lccccc}
\tablenum{2}
\tablecolumns{6}
\tablewidth{0pt}
\tablecaption{Equivalent widths of absorption lines in 
I Zw 18C\tablenotemark{a} \label{Tab2}}
\tablehead{
Line&\multicolumn{2}{c}{region C}&&\multicolumn{2}{c}{region E} \\ \cline{2-3}
\cline{5-6}
    &measured&corrected&&measured&corrected
}
\startdata
H$\delta$&4.9$\pm$0.3&5.8$\pm$0.3&&5.2$\pm$0.3&5.5$\pm$0.3 \\
H$\gamma$& \nodata\tablenotemark{b}& \nodata&&4.0$\pm$0.3&4.7$\pm$0.3 \\
\enddata
\tablenotetext{a}{in \AA.}
\tablenotetext{b}{strongly contaminated by emission.}
\end{deluxetable}

\clearpage

\begin{deluxetable}{lrrcrrrrcrrr}
\tablenum{3}
\scriptsize
\tablecolumns{12}
\tablewidth{0pt}
\tablecaption{Integrated photometric properties of I\ Zw\ 18C \label{Tab3}}
\tablehead{
 &&&&\multicolumn{8}{c}{Models} \\ \cline{5-12}
\multicolumn{1}{c}{Param.} &\multicolumn{2}{c}{Observations\tablenotemark{a}}
&&\multicolumn{4}{c}{Geneva\tablenotemark{b}}
&&\multicolumn{3}{c}{Padua\tablenotemark{c}} \\ \cline{2-3} \cline{5-8} \cline{10-12}          
 &$C$(H$\beta$)=0&=0.3&& 15Myr & 20Myr & 40Myr & 100Myr&& 12.5Myr & 15.8Myr & 100Myr \\ \hline
\multicolumn{1}{c}{(1)}&\multicolumn{1}{c}{(2)}&\multicolumn{1}{c}{(3)}&&
\multicolumn{1}{c}{(4)}&\multicolumn{1}{c}{(5)}&\multicolumn{1}{c}{(6)}&
\multicolumn{1}{c}{(7)}&&\multicolumn{1}{c}{(8)}&\multicolumn{1}{c}{(9)}&
\multicolumn{1}{c}{(10)}
}
\startdata
$V$\tablenotemark{d}     &  19.20&&&&& \\
$U-B$\tablenotemark{e}  
& --0.50&--0.72&&--0.83&--0.77&--0.59&--0.40&&--0.81&--0.75&--0.37 \\
$B-V$\tablenotemark{d}  
&   0.00&--0.20&&--0.17&--0.15&--0.08&--0.01&&--0.18&--0.12&  0.02 \\
$V-I$\tablenotemark{e}  
&   0.10&--0.17&&--0.10&--0.08&  0.04&  0.13&&--0.16&  0.08&  0.21 \\
$V-K$  
&\nodata&\nodata&&--0.15&--0.10&  0.25&  0.47&&--0.25&  0.29&  0.67 \\
\enddata
\tablenotetext{a}{Magnitudes and colors are corrected for 
interstellar extinction with the $C$(H$\beta$) indicated.}
\tablenotetext{b}{Based on Geneva stellar evolutionary tracks with 
$Z$=$Z_\odot$/50 (Lejeune \& Schaerer 2001).}
\tablenotetext{c}{Based on Padua stellar evolutionary tracks with 
$Z$=$Z_\odot$/50 (Tantalo et al. 1996).}
\tablenotetext{d}{Dufour et al. (1996b).}
\tablenotetext{e}{van Zee et al. (1998).}
\end{deluxetable}

\clearpage

%
\begin{deluxetable}{lccr}
\tablenum{4}
\tablecolumns{4}
\tablewidth{0pt}
\tablecaption{Emission line intensities in the H$\alpha$ arc \label{Tab4}}
\tablehead{
Ion   &$F$($\lambda$)/$F$(H$\beta$)&$I$($\lambda$)/$I$(H$\beta$)
&$EW$\tablenotemark{a} }
\startdata
 3727\ [O {\sc ii}]        &0.361$\pm$0.038& 0.374$\pm$0.040&53 \\
 3868\ [Ne {\sc iii}]      &0.139$\pm$0.040& 0.143$\pm$0.042&20 \\
 3889\ He {\sc i} + H8     &0.178$\pm$0.026& 0.196$\pm$0.034&31 \\
 3968\ [Ne {\sc iii}] + H7 &0.212$\pm$0.026& 0.230$\pm$0.034&39 \\
 4101\ H$\delta$           &0.254$\pm$0.028& 0.271$\pm$0.034&49 \\
 4340\ H$\gamma$           &0.455$\pm$0.036& 0.470$\pm$0.041&106 \\
 4363\ [O {\sc iii}]       &0.040$\pm$0.016& 0.040$\pm$0.016&13 \\
 4471\ He {\sc i}          &0.034$\pm$0.014& 0.034$\pm$0.014&9 \\
 4686\ He {\sc ii}         &0.019$\pm$0.011& 0.019$\pm$0.011&7 \\
 4861\ H$\beta$            &1.000$\pm$0.060& 1.000$\pm$0.061&292 \\
 4959\ [O {\sc iii}]       &0.460$\pm$0.036& 0.454$\pm$0.036&179 \\
 5007\ [O {\sc iii}]       &1.417$\pm$0.079& 1.399$\pm$0.079&551 \\
 5876\ He {\sc i}          &0.069$\pm$0.014& 0.067$\pm$0.014&40 \\
 6563\ H$\alpha$           &2.902$\pm$0.140& 2.755$\pm$0.146&1683 \\
 6678\ He {\sc i}          &0.038$\pm$0.012& 0.036$\pm$0.012&21 \\
 6717\ [S {\sc ii}]        &0.051$\pm$0.012& 0.048$\pm$0.011&25 \\
 6731\ [S {\sc ii}]        &0.023$\pm$0.009& 0.022$\pm$0.008&15 \\ \\
 $C$(H$\beta$)       &\multicolumn {3}{c}{0.060$\pm$0.063} \\
 $F$(H$\beta$)\tablenotemark{b} &\multicolumn {3}{c}{ 0.12$\pm$0.01} \\
 $EW$(abs)\ \AA      &\multicolumn {3}{c}{2.3$\pm$3.2}
\enddata
\tablenotetext{a}{in \AA.}
\tablenotetext{b}{in units of 10$^{-14}$ erg\ s$^{-1}$cm$^{-2}$.}
\end{deluxetable}

\clearpage

%
\begin{deluxetable}{lc}
\tablenum{5}
\tablecolumns{2}
\tablewidth{0pt}
\tablecaption{Heavy element abundances in the H$\alpha$ arc \label{Tab5}}
\tablehead{
Parameter&Value }
\startdata
$T_e$(O {\sc iii})(K)                     &18200$\pm$4000     \\
$T_e$(O {\sc ii})(K)                      &15100$\pm$3200     \\
$T_e$(S {\sc iii})(K)                     &16800$\pm$3400     \\
$N_e$(S {\sc ii})(cm$^{-3}$)              &    10$\pm$10        \\ \\
O$^+$/H$^+$($\times$10$^5$)         &0.316$\pm$0.174      \\
O$^{++}$/H$^+$($\times$10$^5$)      &0.964$\pm$0.504      \\
O$^{+3}$/H$^+$($\times$10$^5$)      &0.020$\pm$0.016      \\
O/H($\times$10$^5$)                 &1.301$\pm$0.533      \\
12 + log(O/H)                       &7.114$\pm$0.178      \\ \\
Ne$^{++}$/H$^+$($\times$10$^5$)     &0.206$\pm$0.125      \\
ICF(Ne)                             &1.35\,~~~~~~~~~~     \\
log(Ne/O)                           &--0.670$\pm$0.365~~
\enddata
\end{deluxetable}

\clearpage

%
\begin{deluxetable}{lrcr}
\tablenum{6}
\tablecolumns{4}
\tablewidth{0pt}
\tablecaption{Parameters of the emission lines in the loop \label{Tab6}}
\tablehead{
Line&\multicolumn{1}{c}{$F$\tablenotemark{a}}&\multicolumn{1}{c}{$F/F$(H$\beta$)}
&\multicolumn{1}{c}{$EW$\tablenotemark{b}} }
\startdata
3889 H$\epsilon$+He {\sc i}& 0.73$\pm$0.29&0.183&  54 \\ 
4101 H$\delta$             & 0.91$\pm$0.31&0.228&  87 \\ 
4340 H$\gamma$             & 1.56$\pm$0.38&0.392& 125 \\ 
4861 H$\beta$              & 3.98$\pm$0.96&1.000& 471 \\ 
4959 [O {\sc iii}]         & 0.59$\pm$0.25&0.148&  69 \\ 
5007 [O {\sc iii}]         & 1.94$\pm$0.42&0.487& 229 \\ 
6563 H$\alpha$             &11.55$\pm$0.89&2.902&1396  
\enddata
\tablenotetext{a}{in units of 10$^{-16}$ erg s$^{-1}$cm$^{-2}$.}
\tablenotetext{b}{in \AA.}
\end{deluxetable}

\clearpage

%
%

\clearpage

%

\clearpage

%
%

\clearpage

%
%

\clearpage

%
%

\clearpage

%
%

\clearpage

%
%

\clearpage

%

\clearpage

%
%

\clearpage

%
%

\clearpage

%

\end{document}